\documentclass[aps,pra,preprint]{revtex4-2} 
\usepackage{graphicx,bm,hyperref}
\usepackage{amsmath}
\usepackage{mathrsfs}

\usepackage{hyperref}
\hypersetup{
    colorlinks=true,
    linkcolor=blue,
    citecolor=black,
    filecolor=black,
    urlcolor=black}  

\renewcommand*\eqref[1]{\hyperref[{#1}]{\textup{\normalfont(\ref*{#1})}}}

\newcommand{\upperRomannumeral}[1]{\uppercase\expandafter{\romannumeral#1}}

\usepackage{caption}
\usepackage{subcaption}

\begin{document}

\title{Rotating wave approximation and renormalized perturbation theory}

\author{Peng Wang}
\affiliation{Department of Applied Physics, Yale University, New Haven, CT 06511, USA} 
\email{peng.wang.pw452@yale.edu}

\author{Erik Orvehed Hiltunen}
\affiliation{Department of Mathematics, Yale University, New Haven, CT 06511, USA} 
\email{erik.hiltunen@yale.edu}

\author{John C. Schotland}
\affiliation{Department of Mathematics and Department of Physics, Yale University, New Haven, CT 06511, USA} 
\email{john.schotland@yale.edu}

\begin{abstract}
The rotating wave approximation (RWA) plays a central role in the quantum dynamics of two-level systems.
We derive corrections to the RWA using the renormalization group approach to asymptotic analysis. We study both the Rabi and Jaynes-Cummings models and compare our analytical results with numerical calculations.
\end{abstract}

\maketitle

\newpage

\section{Introduction}

The rotating wave approximation (RWA) plays a central role in the quantum dynamics of two-level systems coupled to a driving field~\cite{Bloch_1940, Shirley_1965, Evans_1968, Cohen-Tannoudji_1973, Seke_1991, Crisp_1991, Ford_1997, Forn-Diaz_2010, Rao_2017, Nalbach_2018, Zeuch_2020, Ermann_2020}. In quantum optics, the RWA arises in both semiclassical and fully quantum theories of light-matter interactions~\cite{Mandel-Wolf,Allen-Eberly}. In either setting, the RWA consists of neglecting counter-rotating terms in the system Hamiltonian. This approximation holds near resonance and when the field is weak. For the Rabi model, where a two-level atom is driven by a classical field, these conditions mean that $\Omega_R \ll \omega$, where $\Omega_R$ is the Rabi frequency, which is proportional to the field strength, and $\omega$ is the frequency of the field.  Alternatively, for the Jaynes-Cummings model, where the atom is coupled to a single-mode quantized field, the necessary conditions correspond to $\Omega_{JC} \ll \omega$, where $\Omega_{JC}$ is proportional to the coupling strength and $\omega$ is the photon frequency. The appeal of the RWA is that it leads to a simple Hamiltonian, with corresponding equations of motion that are readily solved. The RWA is a good approximation at short times, and breaks down at times that are large in comparison to $1/\Omega_R$ or $1/\Omega_{JC}$. 

In this paper, we make use of modern asymptotic analysis to derive corrections to the RWA. The key observation is that the presence of counter rotating terms indicates the existence of multiple time scales. We find that perturbation theory in a suitable small parameter, corresponding to either $\Omega_R/\omega$ or $\Omega_{JC}/\omega$, diverges at long times due to the appearance of secular terms. Such terms are well known in classical mechanics, for instance in the theory of anharmonic oscillators, where they lead to unbounded trajectories. It is well known that this problem can be overcome by means of two-scale asymptotic analysis~\cite{Nayfeh}. We find that while this technique can be used to calculate corrections to the RWA, secular terms still arise at higher order in the asymptotic expansion. A powerful alternative to multi-scale asymptotics borrows ideas from the field-theoretic renormalization group (RG)~\cite{Goldenfeld_1989,Chen_1994,Chen_1996,Kirkinis_2008_1,Kirkinis_2008_2,Kirkinis_2012}. Making use of this approach, we obtain corrections to the RWA that are free of secular terms to finite order in perturbation theory, for both the Rabi and Jaynes-Cummings models. Our results are confirmed by numerical computations. Further numerical studies that go beyond the RWA for the Rabi and Jaynes-Cummings models are reported in ~\cite{Bishop_1996,Irish_2007,Liu_2009,Casanova_2010,Hausinger_2010,He_2012,Nakamura_2021}.

This paper is organized as follows. In Sec.~II we introduce the Rabi model and obtain the equations of motion for the probability amplitudes of the atomic ground and excited states. We then introduce the RWA in this setting. Sec.~III continues the study of the Rabi model first within single-scale perturbation theory, and later by means of a two-scale asymptotic expansion in which secular terms appear. Finally, the RG approach is used to obtain a renormalized multi-scale expansion in which the secular terms are removed. The Jaynes-Cummings model is discussed in Sec.~IV. Following along the same lines as in Sec.~III, we once again obtain a renormalized multi-scale expansion which provides corrections to the RWA at long times and is free of secular terms.  The paper concludes with a discussion in Sec.~V. An alternative approach to the development in Sec.~II is presented in the Appendix.

\section{Rabi model}
\label{sec:Rabi}
The Rabi model describes the interaction of a two-level atom with a classical electromagnetic field. It is the simplest setting in which the RWA arises. In this section, we obtain corrections to the RWA using multi-scale asymptotic analysis. We renormalize the resulting asymptotic series,  removing divergences that are associated with secular terms in the expansion.

\subsection{Equations of motion}
We consider a two-level atom coupled to a classical field. The total Hamiltonian of the system is of the form
\begin{eqnarray}
    \hat{H}=\hat{H}_A+\hat{V}(t) .
\end{eqnarray}
Here the atomic Hamiltonian $\hat{H}_{A}$ is given by
\begin{eqnarray}
\begin{aligned}
    \hat{H}_A&=E_0|0\rangle \langle 0|+E_1|1\rangle \langle 1|\\
    &=E_0 \hat{{1}} +\hbar \omega_0 |1\rangle \langle 1|,
  \end{aligned}\label{eq:hamiltonian_atom}
\end{eqnarray}
where $|0\rangle$ and $|1\rangle$ denote the ground and excited states of the atom, with corresponding energies $E_0$ and $E_1$, respectively and $\hat{{1}}$ denotes the identity operator. 
In addition, we define $\hbar \omega_0 = E_1-E_0$ to be the energy difference between the  ground and excited states. The  time-dependent potential $\hat{V}(t)$ accounts for the interaction between the atom and the field. As is customary, we suppose that the size of the atom is small compared to the wavelength of the electromagnetic field. Accordingly, we treat the atom as an electric dipole and assume that the electric field is spatially uniform.
The potential $\hat{V}$ is thus taken to be 
\begin{eqnarray}
 \hat{V}=-\bm{\hat{d}} \cdot \bm{E} ,
\end{eqnarray}
where $\bm{\hat{d}}$ is the dipole moment operator and $\bm E$ is the electric field. The electric field is monochromatic with frequency $\omega$ and is given by $\bm{E} = \bm{E_0}\cos \omega t$, where $\bm E_0$ is constant. It follows that the potential can be expressed as
\begin{eqnarray}
\hat{V}(t)=-\frac{\langle 0|\bm{\hat{d}} \cdot \bm{E}_0| 1\rangle}{2}(|0\rangle \langle 1|+|1\rangle \langle 0|)(e^{i\omega t}+e^{-i\omega t}), \label{eq:hamiltonian_potential} \ .
\end{eqnarray}
Here $\bm{\hat{d}}$ has been expressed in the atomic basis and is assumed to have odd parity. That is,
$\langle 0|\bm{\hat{d}}|0\rangle=\langle 1|\bm{\hat{d}}|1\rangle=0$. We also assume that the relative phase between the states $|0\rangle$ and $|1\rangle$ can be chosen so that that the quantity $\langle 0|\bm{\hat{d}} \cdot \bm{E}_0|1\rangle$ is real valued.

The dynamics of the system is governed by the Schr\"{o}dinger equation. In the interaction picture, the state
$|\psi_I\rangle$ obeys  
\begin{eqnarray}
i\hbar\frac{d|\psi_{I}(t)\rangle}{dt}=\hat{H}_{I}|\psi_{I}(t)\rangle .
\label{eq:eq_Schrodinger}
\end{eqnarray}
Here the  Hamiltonian $\hat H_I$ is given by
\begin{eqnarray}
\begin{aligned}
    \hat{H}_{I}(t) 
    &=e^{i\hat{H}_A t/\hbar}\hat{V}(t)e^{-i\hat{H}_A t/\hbar}\\
    &=\hbar \Omega_{R}(|0\rangle \langle 1|+|1\rangle \langle 0|)(e^{i\omega t}+e^{-i\omega t}) ,
\end{aligned} 
\label{eq:hamiltonian_interaction}
\end{eqnarray}
where the Rabi frequency $\Omega_R$ is defined by
\begin{eqnarray}
    \Omega_{R} = -\frac{\langle 0|\bm{\hat{d}} \cdot \bm{E}_0| 1\rangle}{2\hbar} .
\end{eqnarray}
The state $|\psi_{I}(t)\rangle$ can be expanded in the atomic basis as
\begin{eqnarray}
    |\psi_{I}(t)\rangle={a(t)}|0\rangle + b(t)|1\rangle.\label{eq:state_interaction}
\end{eqnarray}
We note that $a(t)$ and $b(t)$ are the probability amplitudes that the atom is in its ground state and excited state, respectively. Evidently, the conservation of probability is expressed by the relation
\begin{eqnarray}
|a(t)|^2 + |b(t)|^2 = 1 .
\end{eqnarray}
Making use of  Eqs.\eqref{eq:eq_Schrodinger} and \eqref{eq:hamiltonian_interaction}, we find that the coefficients $a(t)$ and $b(t)$ obey the equations of motion
\begin{eqnarray}
    \nonumber
      i\dot{a}(t)&=&\Omega_R(e^{-i (\omega_0-\omega) t}+e^{-i(\omega_0 +\omega )t})b(t), \\
      i\dot{b}(t)&=&\Omega_R(e^{i(\omega_0 -\omega) t}+e^{i(\omega_0+ \omega) t})a(t) ,
    \label{eq:eq_Rabi_coeff}
\end{eqnarray}
where a dot above a symbol indicates differentiation with respect to time.

\subsection{Rotating wave approximation}
\label{sec:RWA}
Eq.~\eqref{eq:eq_Rabi_coeff} cannot be directly integrated. However, a solution can be obtained by introducing the rotating wave approximation (RWA). To this end, it will prove useful to write Eq.~\eqref{eq:eq_Rabi_coeff} in dimensionless form by rescaling the time $t$ by $\Omega_R t$. We thus obtain
\begin{eqnarray}
    \begin{aligned}
      i\dot{a}(t) &= (e^{-i\delta_R t}+e^{-i\Delta_R t})b(t), \\
      i\dot{b}(t) &= (e^{i\delta_R t}+e^{i\Delta_R t})a(t),
      \label{eq:eq_Rabi}
     \end{aligned}       
\end{eqnarray}
where 
\begin{eqnarray}
    \delta_R = \frac{\omega_0-\omega}{\Omega_R}, \quad  \Delta_R = \frac{\omega_0+\omega}{\Omega_R}.
\end{eqnarray}
We note that $\Omega_R \ll \omega$ at optical frequencies. Near resonance, where $\omega \approx \omega_0$, we see that $\Delta_R \gg \delta_R$. In the RWA, we neglect the fast rotating terms  $e^{\pm i\Delta_R t}$ in Eq.~\eqref{eq:eq_Rabi}, which thus becomes
\begin{eqnarray}
    \begin{aligned}
      i\dot{a}(t) &= e^{-i\delta_R t}b(t), \\
      i\dot{b}(t) &= e^{i\delta_R t} a(t) .
      \label{eq:eq_Rabi_rwa}
     \end{aligned}       
\end{eqnarray}
Suppose that the system is on resonance with $\delta_R=0$ and that the atom is initially in the ground state, so that $a(0)=1$ and $b(0)=0$.
The solution to Eq.~\eqref{eq:eq_Rabi_rwa} is then given by
\begin{eqnarray}
    \begin{aligned}
      a(t)&=\cos t, \\
      b(t)&=-i \sin t.
    \end{aligned}       
    \label{eq:rabi_rwa_solution}
\end{eqnarray}
The probability that the atom is in the ground state is $|a(t)|^2 = \cos^2(\Omega_R t)$, which oscillates at twice the Rabi frequency. In Fig.~1 we compare this result with the numerical solution to Eq.~\eqref{eq:eq_Rabi}
with $\Delta_R =50$. We see that at short times, the RWA is quite accurate, but is less accurate at long times, as may be expected.


\begin{figure}[t]
     \centering
     \begin{subfigure}[b]{0.48\textwidth}
         \centering
         \includegraphics[width=\textwidth]{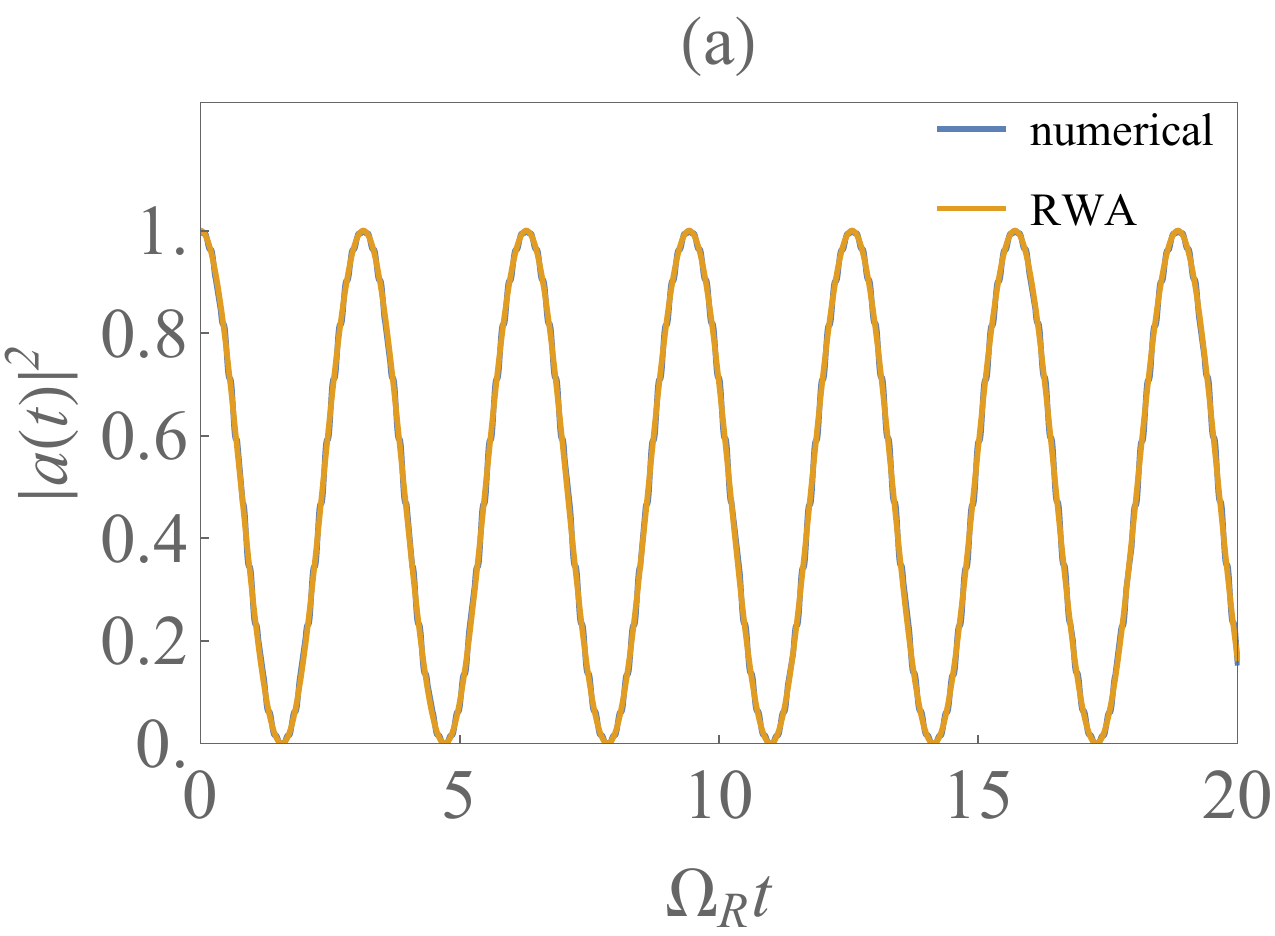}
        
     \end{subfigure}
     \begin{subfigure}[b]{0.48\textwidth}
         \centering
         \includegraphics[width=\textwidth]{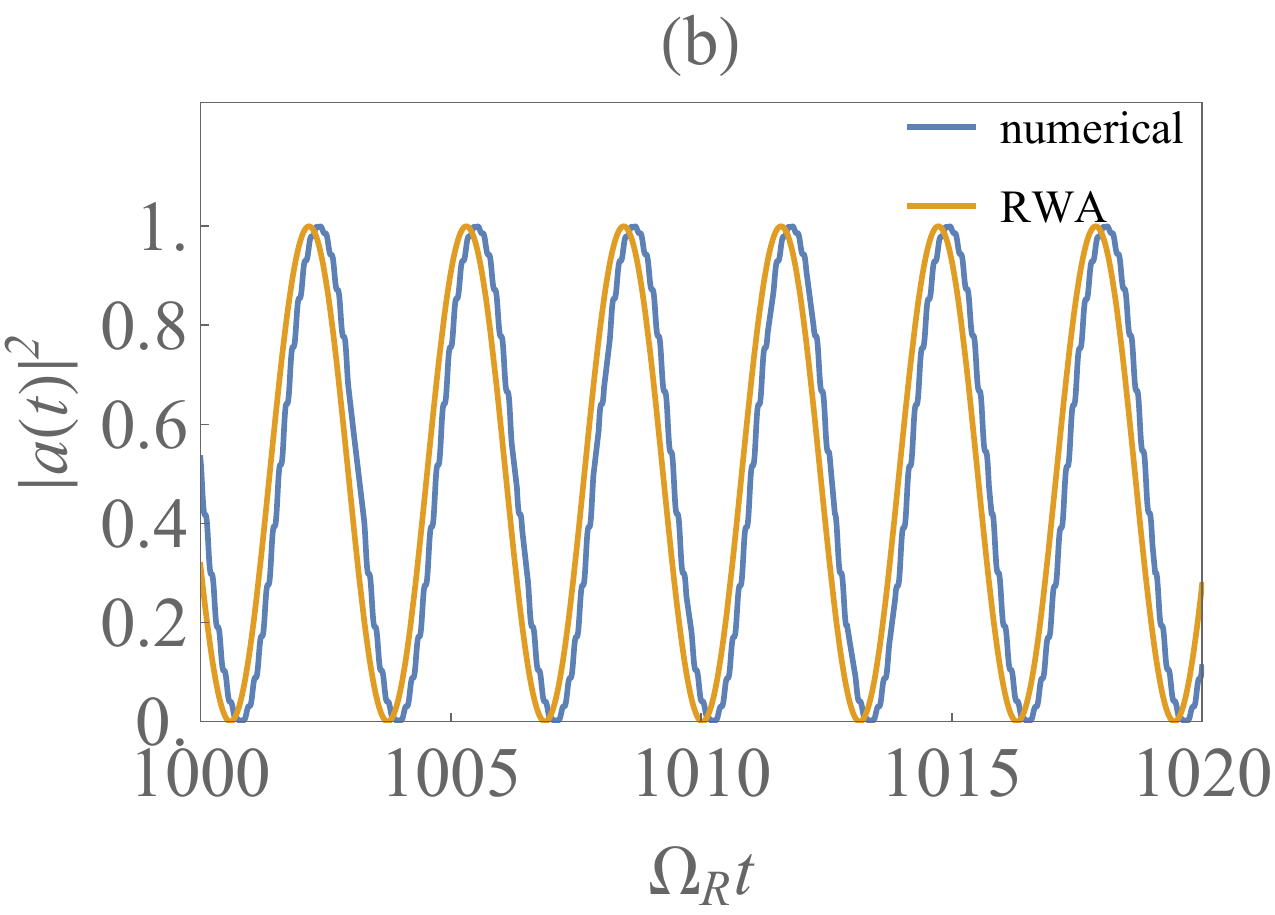}
         
     \end{subfigure}

        \caption{\footnotesize \footnotesize Comparison of the RWA for the Rabi model with numerical solutions to the equations of motion
        Eq.~\eqref{eq:eq_Rabi}. The probability $|a(t)|^2$ is plotted with $\delta_R=0$ and $\Delta_R=50$ for (a) short times and (b) long times.}
     \end{figure}

\section{Renormalized perturbation theory}
\subsection{Single-scale expansion}
We now consider the problem of calculating perturbative corrections to the RWA. In doing so, we account for the effects of the fast rotating terms in Eq.~\eqref{eq:eq_Rabi}. For simplicity, we will focus on the resonant case with $\delta_R=0$. Eq.~\eqref{eq:eq_Rabi} then becomes
\begin{eqnarray}
    \begin{aligned}
      i\dot{a}(t)&=(1+e^{-i\Delta_R t})b(t), \\
      i\dot{b}(t)&=(1+e^{i\Delta_R t})a(t).
    \end{aligned}
    \label{eq:eq_rabi_rwa_resonance}
\end{eqnarray}
We begin by rescaling the time $t$ by $\epsilon t$, where $\epsilon$ is a small parameter. Here we have identified $\epsilon$ with $1/\Delta_R$, so that $\epsilon \ll 1$ when $\Delta_R \gg 1$. 
The rescaled Eq.~\eqref{eq:eq_rabi_rwa_resonance} is of the form
\begin{eqnarray}
    \begin{aligned}
      i\dot{a_\epsilon}(t) &= \epsilon (1+e^{-i t})b_\epsilon(t), \\
      i\dot{b_\epsilon}(t) &= \epsilon (1+e^{i t})a_\epsilon(t),
    \end{aligned}      
    \label{eq:rabi_eq_pertubation}
\end{eqnarray}
where $a_\epsilon(t) = a(\epsilon t)$ and $b_\epsilon(t) = b(\epsilon t)$. 

The solution to Eq.~\eqref{eq:rabi_eq_pertubation} is obtained by expanding $a_\epsilon(t)$ and $b_\epsilon(t)$
in powers of $\epsilon$:
\begin{eqnarray}
    \begin{aligned}
      a_\epsilon(t)&=a_0(t)+\epsilon a_1(t)+\epsilon ^2 a_2(t)+\cdots,\\
      b_\epsilon(t)&= b_0(t)+\epsilon b_1(t)+\epsilon ^2 b_2(t)+\cdots.
    \end{aligned}     
    \label{eq:rabi_one_scale_expansion}
\end{eqnarray}
The initial conditions $a(0)=1$ and $b(0)=0$ become $a_\epsilon (0)=1$ and $b_\epsilon(0)=0$, which yields
\begin{eqnarray}
    \begin{cases}
      a_0(0)=1, \\
      a_1(0)=a_2(0)=\cdot \cdot \cdot=0,\\
      b_0(0)= b_1(0)=\cdot \cdot \cdot=0.    
    \end{cases}     
    \label{eq:rabi_one_scale_initial}
\end{eqnarray}
Substituting Eq.~\eqref{eq:rabi_one_scale_expansion} into Eq.~\eqref{eq:rabi_eq_pertubation} and collecting terms of order $O(1)$, $O(\epsilon)$ and  $O(\epsilon^2)$ we find that
\begin{eqnarray}
\begin{aligned}
i\dot{a}_0(t)&=0,\\
i\dot{a}_1(t)&=(1+e^{-it})b_0(t),\\
i\dot{a}_2(t)&=(1+e^{-it})b_1(t).\\
\end{aligned}
\label{eq:rabi_one_scale_3terms}
\end{eqnarray}
The solutions to Eqs.~\eqref{eq:rabi_one_scale_3terms} that obey the initial conditions Eq.~\eqref{eq:rabi_one_scale_initial} are given by
\begin{eqnarray}
    \begin{aligned}
      a_0(t)&=1, \\
      a_1(t)&=0,\\
      a_2(t)&=e^{it}-ite^{-it}-\frac{1}{2}t^{2}-1.\\
    \end{aligned}       
    \label{eq:rabi_one_time_a}
\end{eqnarray}
Following the same procedure, we find that the terms in the asymptotic expansion of 
$b_\epsilon(t)$ are of the form
\begin{eqnarray}
    \begin{aligned}
      b_0(t)&=0, \\
      b_1(t)&=-it-e^{it}+1,\\
      b_2(t)&=0.\\
    \end{aligned}   
    \label{eq:rabi_one_time_b}
\end{eqnarray}

We now make a crucial observation. The expansions we have obtained for $a_\epsilon(t)$ and $b_\epsilon(t)$ contain \emph{secular terms} which diverge at long times. This is illustrated in Fig.~2, 
which shows numerical calculations of the probabilities $|a(t)|^2$ and $|b(t)|^2$ compared with the RWA and second order perturbation theory. It can be seen that perturbation theory break down for times $t = O(1)$.

\begin{figure}[t]
     \centering
     \begin{subfigure}[b]{0.45\textwidth}
         \centering
         \includegraphics[width=\textwidth]{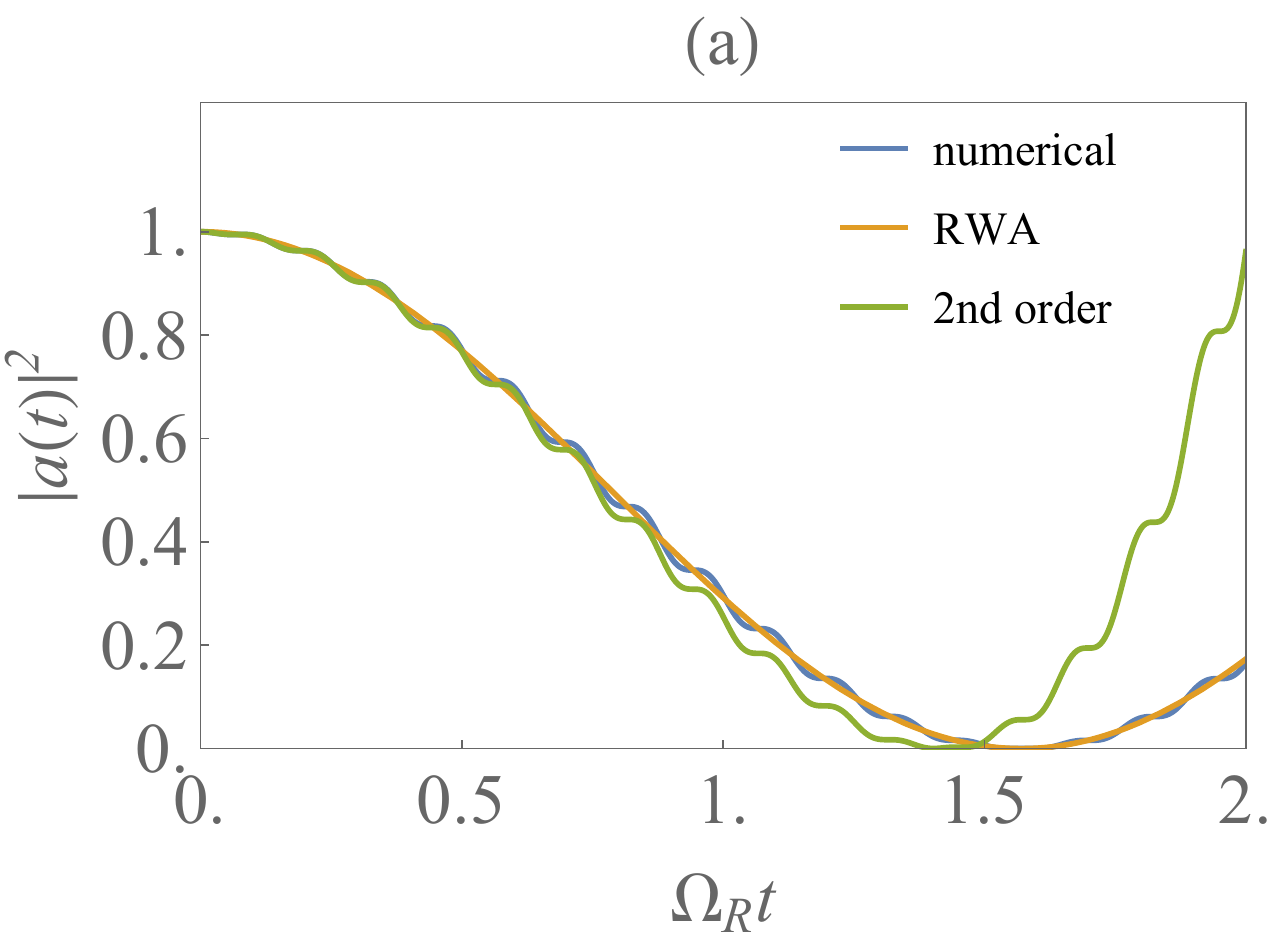}
     \end{subfigure}
     \hfill
     \begin{subfigure}[b]{0.45\textwidth}
         \centering
         \includegraphics[width=\textwidth]{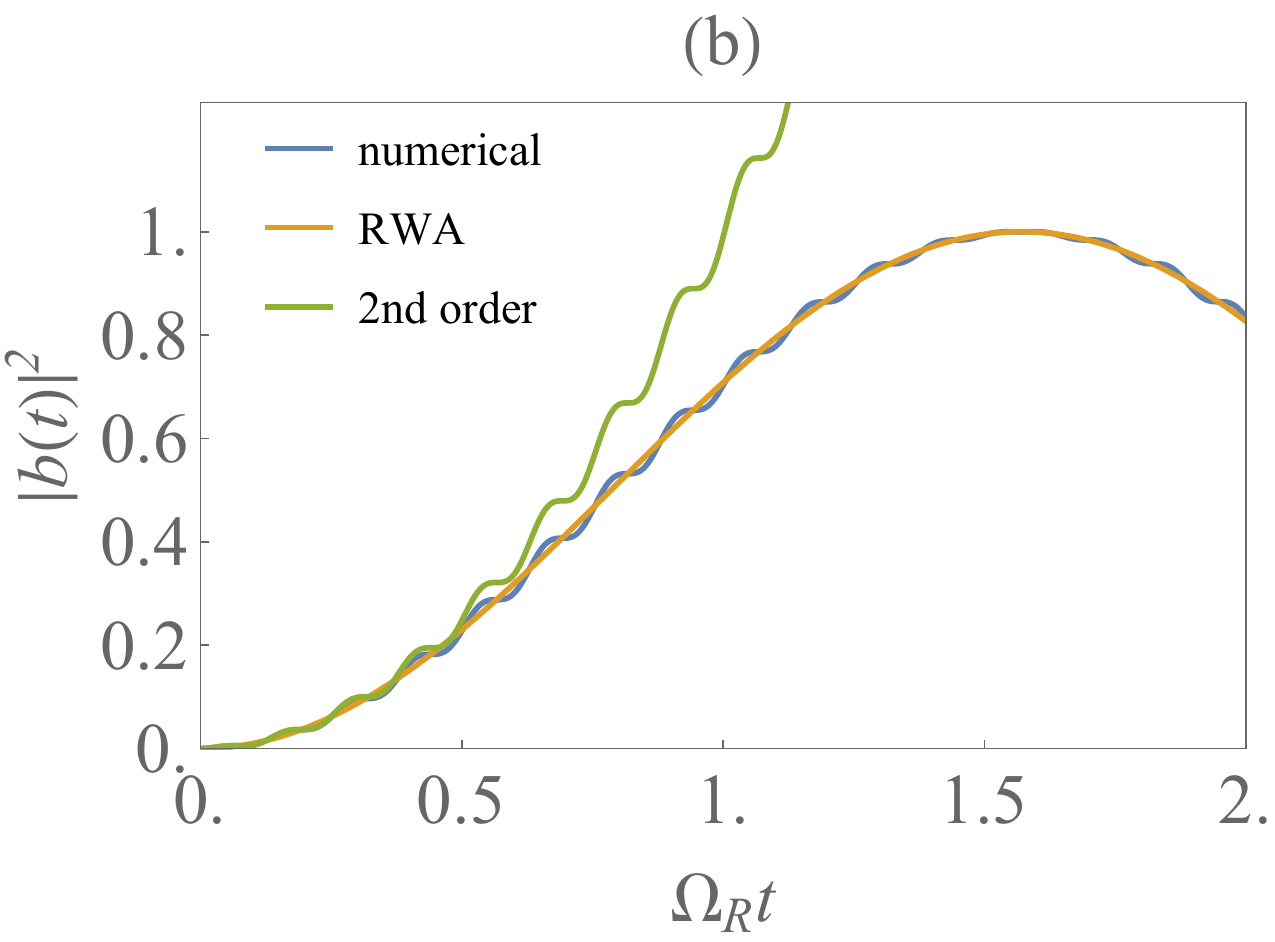}
     \end{subfigure}
    
 \caption{\footnotesize Illustrating the effect of secular terms. The RWA and second order perturbation theory are compared with numerical computations of (a) $|a(t)|^2$ and (b) $|b(t)|^2$ for $\delta_R=0$ and $\Delta_R=50$.}
\end{figure}

\subsection{Multi-scale expansion}
The divergence of the expansion \eqref{eq:rabi_one_scale_expansion} is due to the presence of both fast and slow time scales. 
This problem may be handled by introducing a multi-scale expansion, which has the effect of removing the  secular terms~\cite{Nayfeh}. The expansion is of the form
\begin{eqnarray}
    \begin{aligned}
      a_\epsilon(t_1,t_2)=a_0(t_1,t_2)+\epsilon a_1(t_1,t_2)+\epsilon ^2 a_2(t_1,t_2)+\cdots, \\
      b_\epsilon(t_1,t_2)=b_0(t_1,t_2)+\epsilon b_1(t_1,t_2)+\epsilon ^2 b_2(t_1,t_2)+\cdots.
    \end{aligned}   
    \label{eq:rabi_expansion_two_scale}
\end{eqnarray}
Here we consider $a$ and $b$ to be functions of two independent variables $t_1$ and $t_2$, where $t_1 = t$ and $t_2 = \epsilon t$. We will see that the variable $t_2$ describes the time evolution of the system on scales that are larger than $t_1$.  It follows that 
the time derivative $d/dt$ transforms according to
\begin{eqnarray}
\label{eq:derivative}
  \frac{d}{dt} \rightarrow{\partial_{t_1}+\epsilon \partial_{t_2}}.
\end{eqnarray}
Thus Eq.~\eqref{eq:rabi_eq_pertubation} becomes
\begin{eqnarray}
    \begin{aligned}
      i(\partial_{t_1}+\epsilon \partial_{t_2}){a_\epsilon}(t_1,t_2)=\epsilon (1+e^{-i t})b_\epsilon(t_1,t_2), \\
      i(\partial_{t_1}+\epsilon \partial_{t_2}){b_\epsilon}(t_1,t_2)=\epsilon (1+e^{i t})a_\epsilon(t_1,t_2).
    \end{aligned}  
    \label{eq:rabi_eq_two_scale}
\end{eqnarray}
The initial conditions Eq.~\eqref{eq:rabi_one_scale_initial} then become
\begin{eqnarray}
    \begin{aligned}
      a_0(0,0)=1, \\
      a_1(0,0)=a_2(0,0)=\cdot \cdot \cdot=0,\\
      b_0(0,0)= b_1(0,0)=\cdot \cdot \cdot=0.
    \end{aligned}   
    \label{eq:rabi_two_scale_initial}
\end{eqnarray}
By inserting Eq.~\eqref{eq:rabi_expansion_two_scale} into Eq.~\eqref{eq:rabi_eq_two_scale} and collecting terms of the same order in $\epsilon$, we find that
\begin{eqnarray}
\begin{aligned}
    i\partial_{t_1}a_n(t_1,t_2)=0 , \quad n=0 ,\\
    i\bigl(\partial_{t_1}a_{n}(t_1,t_2)+\partial_{t_2}a_{n-1}(t_1,t_2)\bigr)=(1+e^{-it_1})b_{n-1}(t_1,t_2) , \quad n\geq 1,\\
\end{aligned}  
\label{eq:rabi_two_scale_eqa}
\end{eqnarray}
and 
\begin{eqnarray}
\begin{aligned}
    i\partial_{t_1}b_n(t_1,t_2)=0 , \quad n=0, \\
    i\bigl(\partial_{t_1}b_n(t_1,t_2)+\partial_{t_2}b_{n-1}(t_1,t_2)\bigr)=(1+e^{it_1})a_{n-1}(t_1,t_2) , \quad n\geq 1.
\end{aligned}   
\label{eq:rabi_two_scale_eqb}
\end{eqnarray}

Solving Eqs.~\eqref{eq:rabi_two_scale_eqa} and \eqref{eq:rabi_two_scale_eqb} for $n=0,1$ gives
\begin{eqnarray}
\label{eq:secular_1}
    \begin{aligned}
      a_0(t_1,t_2)&=\alpha_0(t_2), \\
      a_1(t_1,t_2)&=-(i\beta_0+\dot{\alpha_0})t_1-\beta_0e^{-it_1}+\alpha_1(t_2),\\
    \end{aligned}       
\end{eqnarray}
and
\begin{eqnarray}
\label{eq:secular_2}
    \begin{aligned}
      b_0(t_1,t_2)&=\beta_0(t_2), \\
      b_1(t_1,t_2)&=-(i\alpha_0+\dot{\beta_0})t_1-\alpha_0e^{it_1}+\beta_1(t_2).\\  
    \end{aligned}       
\end{eqnarray}
Here $\alpha_{n}(t_2)$ and $\beta_{n}(t_2)$ for $n=0,1$ are independent of $t_1$. Next, we choose  $\alpha_{0}(t_2)$ and $\beta_{0}(t_2)$ so that the secular terms in $t_1$ in Eqs.~\eqref{eq:secular_1} and \eqref{eq:secular_2} are removed. That is, we set
\begin{eqnarray}
    \begin{aligned}
      i\beta_0+\dot{\alpha_0}=0, \\
      i\alpha_0+\dot{\beta_0}=0 .
    \end{aligned}       
\end{eqnarray}
The solution to the above equations that obeys the initial conditions \eqref{eq:rabi_two_scale_initial} is given by
\begin{eqnarray}
    \begin{aligned}
     \alpha_0(t_2) &=\cos t_2, \\
    \beta_0(t_2) &=-i \sin t_2.
    \end{aligned}       
\end{eqnarray}
Thus to zeroth order we have
\begin{eqnarray}
    \begin{aligned}
     a_0(t_2) &=\cos t_2, \\
    b_0(t_2) &=-i \sin t_2 ,
    \end{aligned}       
\end{eqnarray}
which agrees with Eq.~\eqref{eq:rabi_rwa_solution}. We conclude that to lowest order, the multi-scale expansion coincides with the RWA.

The higher order terms in the multi-scale expansion can be obtained by following along similar lines.  We find that at first order, 
\begin{eqnarray}
    \begin{aligned}
    \alpha_1(t_2) &=0, \\
    \beta_1(t_2) &=\cos t_2,
    \end{aligned}       
\end{eqnarray}
and at second order,
\begin{eqnarray}
    \begin{aligned}
    \alpha_2(t_2)=-\cos t_2+\frac{1}{2}t_2 \sin t_2, \\
    \beta_2(t_2)=\frac{i}{2}(t_2\cos t_2+\sin t_2).
    \end{aligned}      
    \label{eq:rabi_two_scale_2ndorder_alpha_beta}
\end{eqnarray}
The corresponding terms in the multi-scale expansion are given by
\begin{eqnarray}
    \begin{aligned}
     a_1(t_1,t_2)&=-i\sin t_2e^{-it_1},\\
     b_1(t_1,t_2)&=\cos t_2-\cos t_2e^{it_1},
    \end{aligned}       
\end{eqnarray}
and
\begin{eqnarray}
    \begin{aligned}
     a_2(t_1,t_2)&=\frac{1}{2}t_2\sin t_2-\cos t_2+\cos t_2e^{it_1},\\
     b_2(t_1,t_2)&=\frac{i}{2}(t_2\cos t_2+\sin t_2)+i\sin t_2(e^{it_1}-e^{-it_1}).
    \end{aligned}       
\end{eqnarray}
Using the above results, we find that $a_\epsilon$ is given by
\begin{eqnarray}
\begin{aligned}
    a_\epsilon(t_1,t_2) =\cos t_2+\epsilon\left(-i\sin t_2e^{-it_1}\right)+\epsilon^2 \left(\frac{1}{2}t_2\sin t_2-\cos t_2+\cos t_2e^{it_1}\right)+O(\epsilon^3).
\end{aligned}
\label{eq:rabi_expansion_two_scale_a2nd}
\end{eqnarray}
Evidently we have eliminated the secular terms to first order in $\epsilon$. However, secular terms reappear at second order. Thus the resulting expansion diverges at times of order $1/\epsilon^2$. It can be seen that this pattern continues at all orders in the multi-scale expansion. 


The multi-scale expansion \eqref{eq:rabi_expansion_two_scale_a2nd} provides corrections to the RWA, as illustrated in Fig.~3. The probability $|a(t)|^2=|a_\epsilon(t/\epsilon)|^2$ is plotted with $\epsilon=1/\Delta_R = 0.1$ and $\delta_R=0$. We also show the error $\bigl||a(t)|^2-|a_{\rm num}(t)|^2\bigr|$, where $a_{\rm num}$ is the numerical solution to Eq.~\eqref{eq:eq_rabi_rwa_resonance}. The effect of the secular terms is evident at long times. 

\begin{figure}[t]
     \centering
     \begin{subfigure}[b]{0.45\textwidth}
         \centering
         \includegraphics[width=\textwidth]{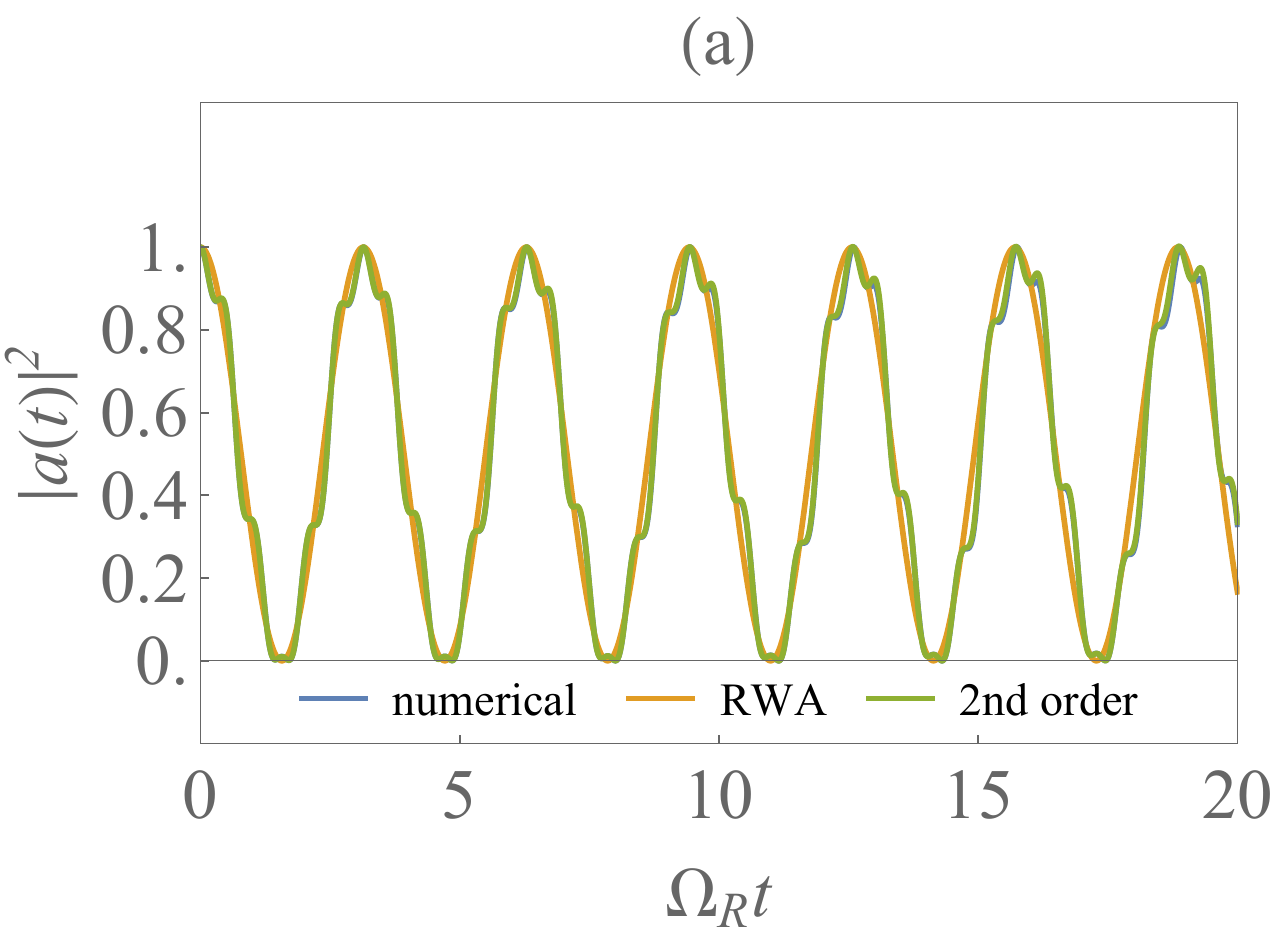}
        
     \end{subfigure}
     \hfill
     \begin{subfigure}[b]{0.45\textwidth}
         \centering
         \includegraphics[width=\textwidth]{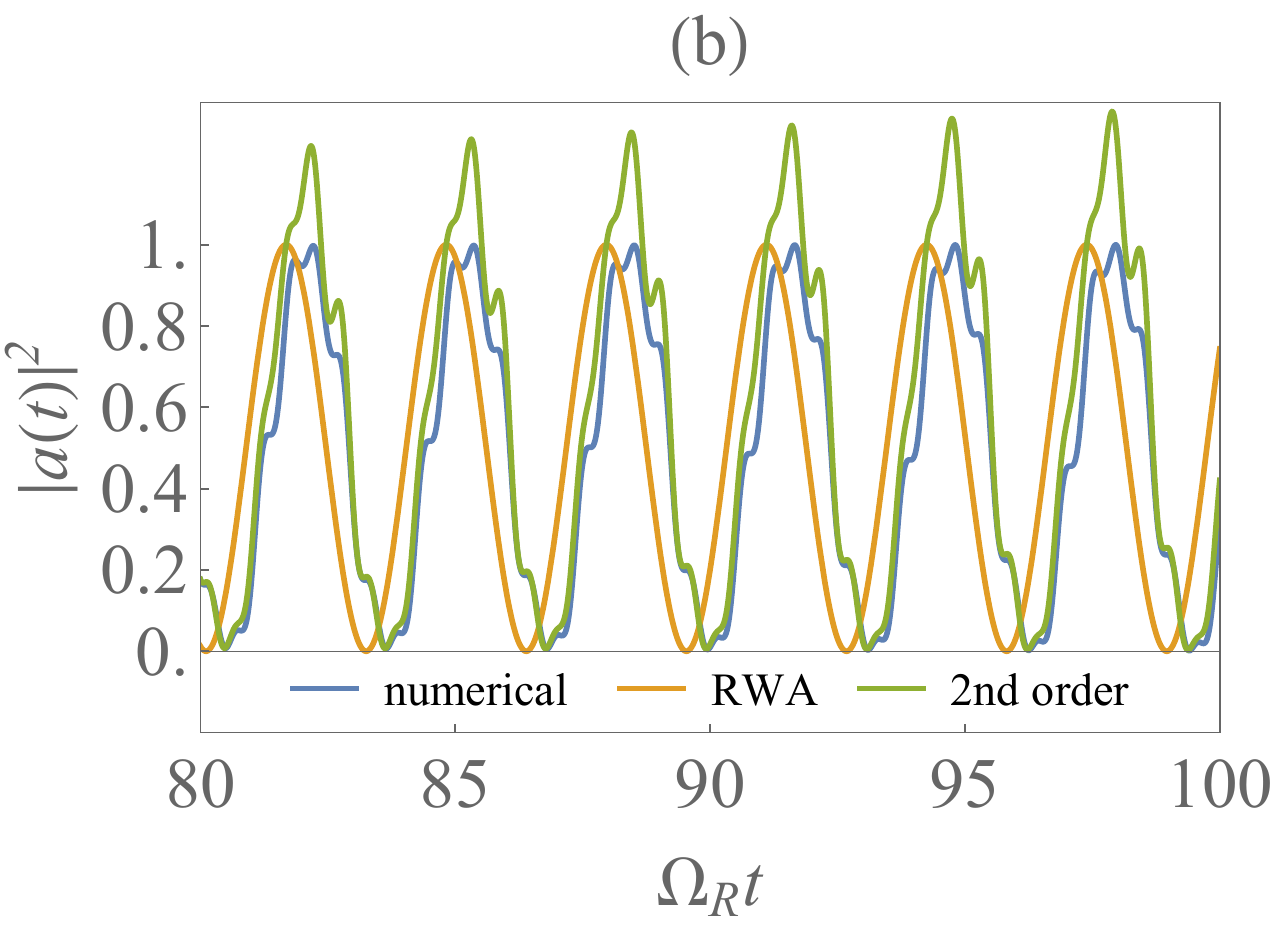}
         
     \end{subfigure}
     
     \begin{subfigure}[b]{0.45\textwidth}
         \centering
         \includegraphics[width=\textwidth]{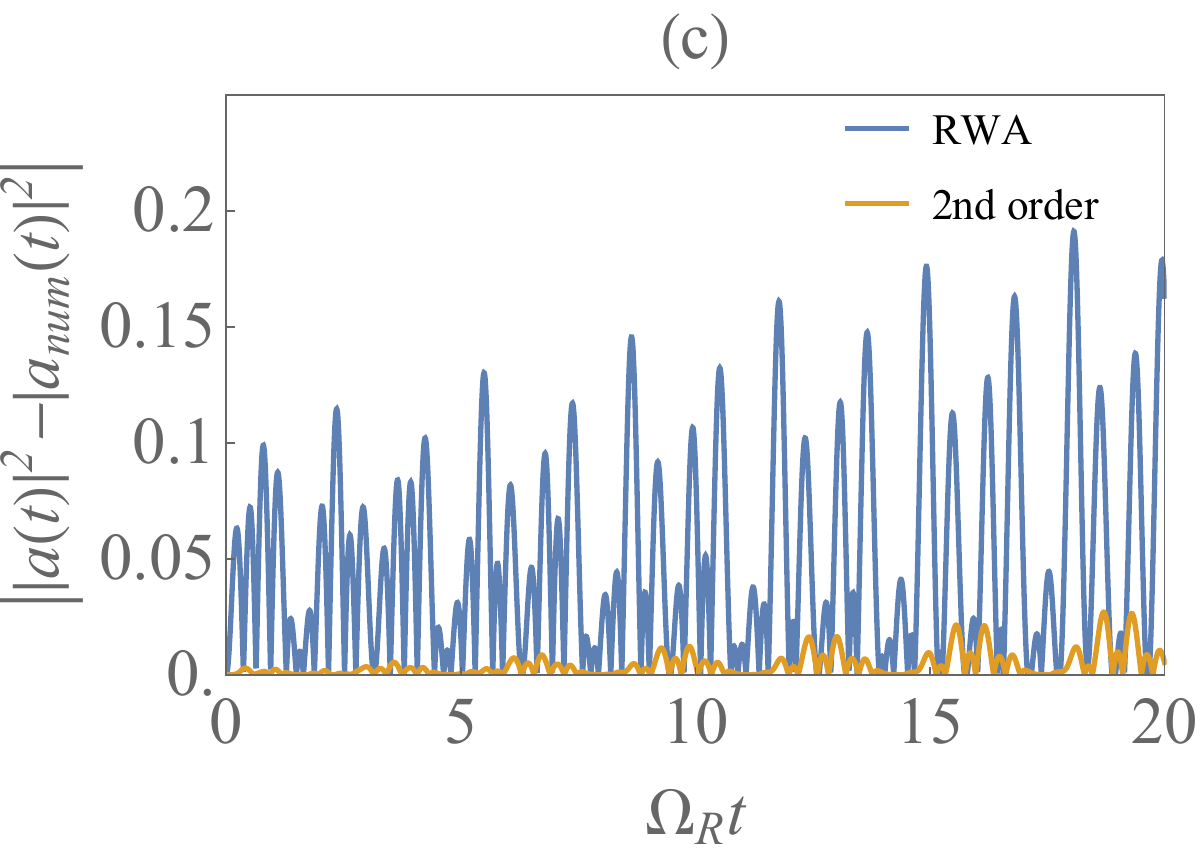}
        
     \end{subfigure}
     \hfill
     \begin{subfigure}[b]{0.45\textwidth}
         \centering
         \includegraphics[width=\textwidth]{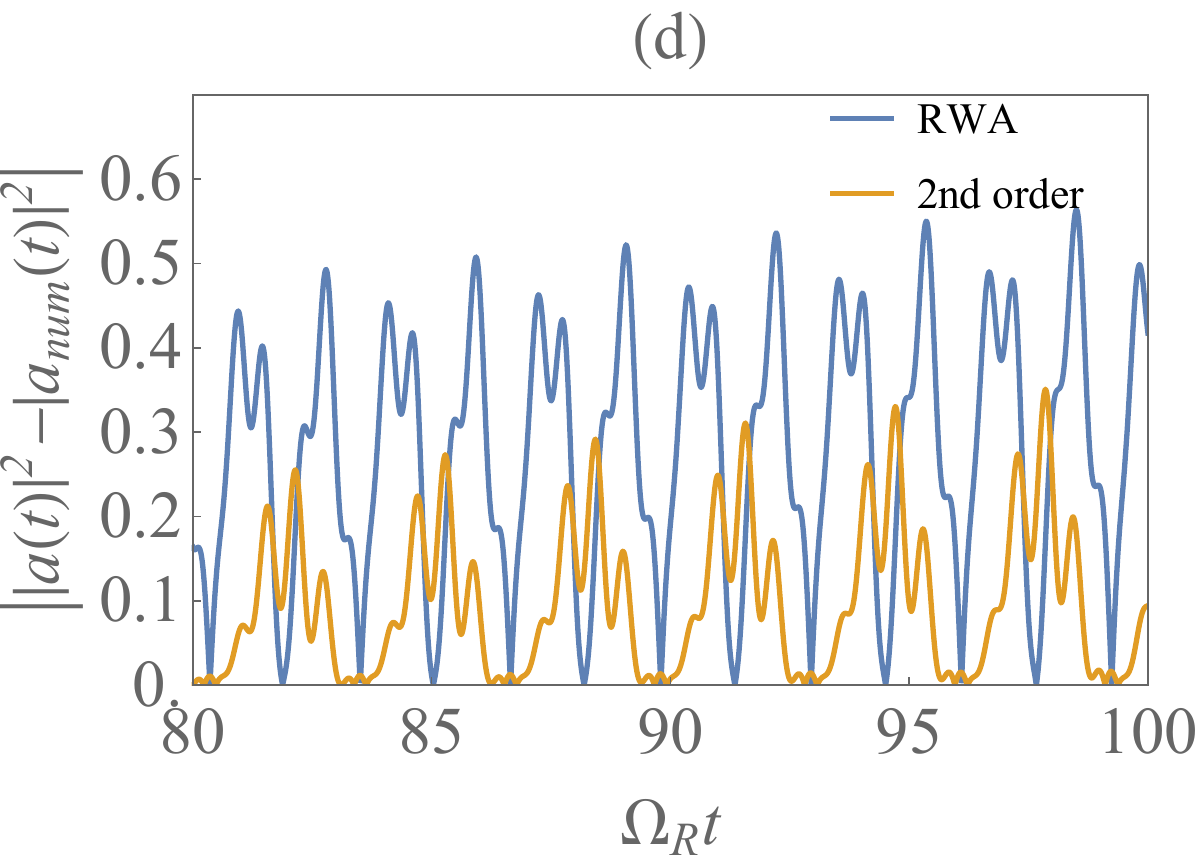}
         
     \end{subfigure}
\caption{\footnotesize Comparison of the RWA for the Rabi model with the two-scale expansion Eq.~\eqref{eq:rabi_expansion_two_scale_a2nd} and numerical solutions to the equations of motion Eq.~\eqref{eq:eq_Rabi}. The probability $|a(t)|^2$ is plotted with $\delta_R=0$ and $\Delta_R=10$ for short times [(a) and (c)] and long times [(b) and (d)]. The second order approximation provides a better result than the RWA at short times, but diverges for times of the order of  $O(1/\epsilon ^{2})$.}
\end{figure}

\subsection{Renormalized multi-scale expansion}

We have seen that the multi-scale expansion \eqref{eq:rabi_expansion_two_scale_a2nd} diverges at long times due to the presence of secular terms. This problem can be overcome by utilizing the renormalization group (RG) method of asymptotic analysis~\cite{Goldenfeld_1989,Chen_1994,Chen_1996,Kirkinis_2008_1,Kirkinis_2008_2,Kirkinis_2012}.
The key idea is to group the secular terms with appropriate non-secular terms so that each group can be renormalized to yield an asymptotic series that neither diverges nor decays to zero at long times. 

We begin by rewriting Eq.~\eqref{eq:rabi_expansion_two_scale_a2nd} so that the secular term in $a_2(t_1,t_2)$ is grouped with the non-secular term $a_0(t_2)$. We thus obtain
\begin{eqnarray}
    a_\epsilon(t_1,t_2)=\cos t_2+\frac{\epsilon^2}{2}t_2 \sin t_2 - i\epsilon \sin t_2e^{-it_1}+\epsilon^2(-\cos t_2+\cos t_2e^{it_1})+O(\epsilon^3).
 \label{eq:rabi_rg_rearrange}
\end{eqnarray}
Next, we seek a suitable function $A_{\epsilon}(t_2)$ whose expansion in powers of $\epsilon$ coincides with the first two terms in Eq.~\eqref{eq:rabi_rg_rearrange}. That is, $A_{\epsilon}$ is constructed so that
\begin{eqnarray}
\begin{aligned}
A_\epsilon(t_2) &= A_0+\epsilon \partial_\epsilon A_\epsilon (0) +\frac{\epsilon ^2}{2} \partial^2_{\epsilon}A_\epsilon (0) +O(\epsilon ^3)\\
    &=\cos t_2+\frac{\epsilon ^2}{2}t_2 \sin t_2+O(\epsilon ^3)\\
    &=  A^{+}_\epsilon(t_2)+A^{-}_\epsilon (t_2)+O(\epsilon^3) .
\end{aligned}\label{eq:rabi_rg_function_A}
\end{eqnarray}
Here the functions $A_\epsilon^{\pm} (t_2)$ are defined by
\begin{eqnarray}
A_\epsilon ^{\pm} (t_2) = \frac{1}{2}e^{\pm it_2}\left(1 \mp \frac{i}{2}\epsilon ^2 t_2+O(\epsilon ^3)\right) ,\label{eq:rabi_rg_A_expansion}
\end{eqnarray}
which we rewrite as
\begin{eqnarray}
\label{eq:Apm}
    A_{\epsilon} ^{\pm}(t_2)=A_0^{\pm}\left(1+\epsilon y^{\pm}_1+\epsilon ^2 y^{\pm}_2 +O(\epsilon ^3)\right) ,
\end{eqnarray}
where
\begin{eqnarray}
   A_0^{\pm}=\frac{1}{2}e^{\pm i t_2}, \hspace{0.2cm} y_1^{\pm}(t_2)=0, \hspace{0.2cm} y_2^{\pm}(t_2)=\mp\frac{i}{2}t_2.
\end{eqnarray}
We observe that the terms $y_2^{\pm}(t_2)$ are the secular terms that are to be removed. To proceed, we note that $A_\epsilon^{\pm}(t_2)$ obeys the ordinary differential equation
\begin{eqnarray}
\begin{aligned}
    \frac{dA_\epsilon ^{\pm} (t_2)}{dt_2}&=A_0 ^{\pm} \left(\epsilon \frac{dy_1^{\pm}}{dt_2}+\epsilon ^2 \frac{dy_2^{\pm}}{dt_2}+O(\epsilon ^3)\right)\\
    &=\frac{A_\epsilon ^{\pm} (t_2)}{1+\epsilon y^{\pm}_1+\epsilon ^2 y^{\pm}_2 +O(\epsilon ^3)}\left(\epsilon \frac{dy_1^{\pm}}{dt_2}+\epsilon ^2 \frac{dy_2^{\pm}}{dt_2}+O(\epsilon ^3)\right)\\
    &=A_\epsilon^{\pm}(t_2) \left[\epsilon \frac{dy_1^{\pm}}{dt_2}+\epsilon^2 \frac{d}{dt_2}\left(y_2^\pm
    -\frac{1}{2}{y_1^\pm}^2\right)+O(\epsilon^3)\right] .
\end{aligned}\label{eq:rabi_rg_diffeq_A}
\end{eqnarray}
Here we have used Eq.~\eqref{eq:Apm} and have expanded the denominator in the second line in powers of $\epsilon$. The solution to the resulting differential equation is given by
\begin{eqnarray}
\begin{aligned}
A_\epsilon ^{\pm}(t_2)&=A_\epsilon ^{\pm}(0) e^{\mp i \frac{\epsilon ^2}{2} t_2} + O(\epsilon^3) \\
&=\frac{1}{2}\exp\left[{\pm it_2\left(1-\frac{\epsilon ^2}{2}\right)}\right] + O(\epsilon^3) .
\end{aligned}
\label{eq:rabi_rg_A_solution}
\end{eqnarray}
We note that the renormalized function $A_{\epsilon}^{\pm}(t_2)$ no longer contains secular terms. Using this result along with Eq.~\eqref{eq:rabi_rg_rearrange}, we find that the renormalized perturbation expansion for 
$a_\epsilon (t_1,t_2)$  is given by
\begin{eqnarray}
\label{eq:RWAfinal}
    a_\epsilon ^R(t_1,t_2)=\cos\left(1-\frac{\epsilon ^2}{2}\right)t_2-i\epsilon \sin t_2e^{-it_1}+\epsilon^2(-\cos t_2+\cos t_2e^{it_1})+O(\epsilon^3).
\end{eqnarray}
Evidently, the above result provides corrections to the RWA which are finite at long times. 

The probability $|a^R(t)|^2=|a_\epsilon ^R (t/\epsilon)|^2$ is plotted in Fig.~4  with $\epsilon=1/\Delta_R = 0.1$. Also shown is the comparison with Eq.~\eqref{eq:rabi_expansion_two_scale_a2nd}. 
We see that the divergence at time $O(1/\epsilon ^{2})$ has been suppressed by renormalization.

\begin{figure}[t]
     \centering
     \begin{subfigure}[b]{0.45\textwidth}
         \centering
         \includegraphics[width=\textwidth]{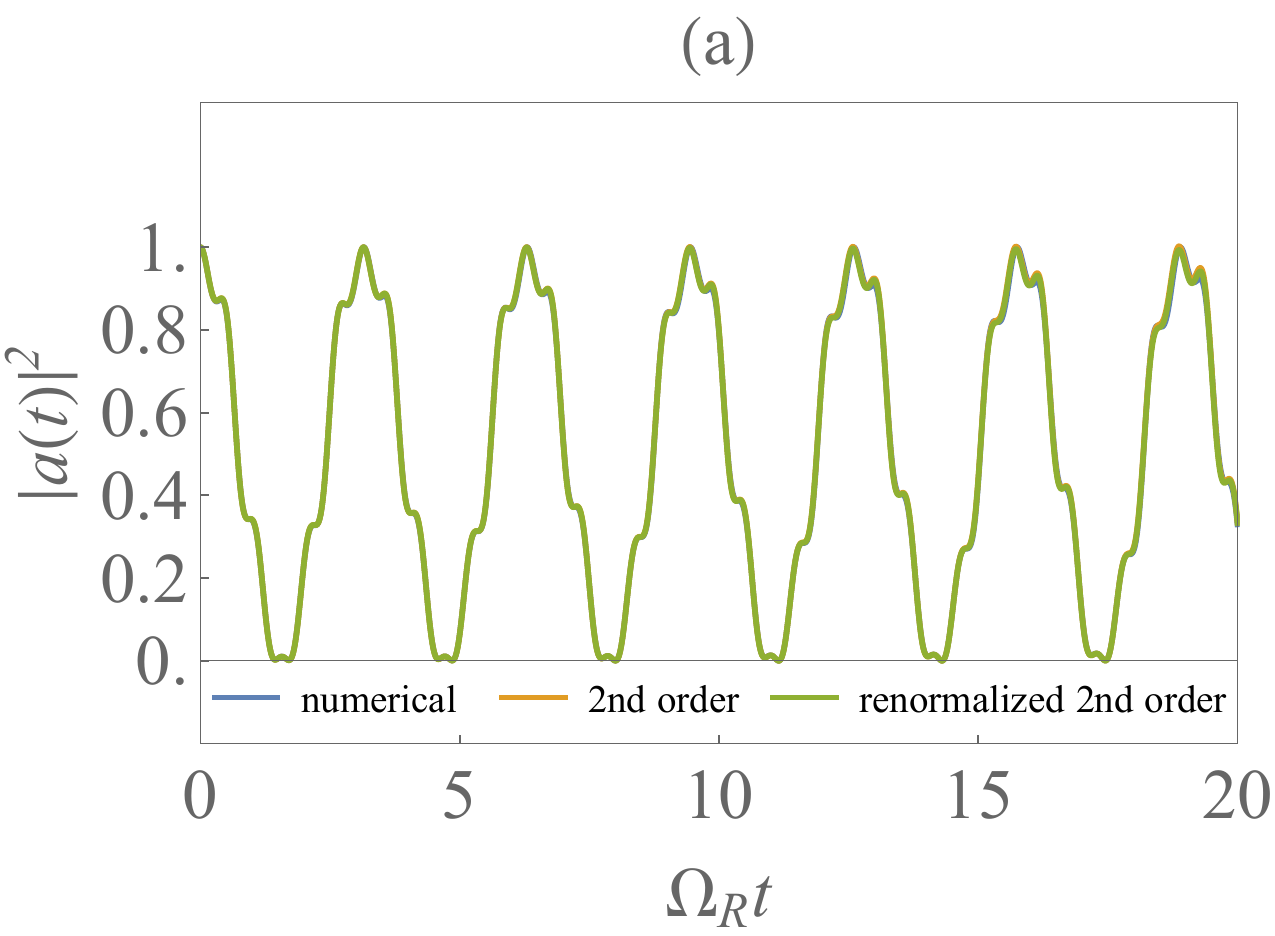}
        
     \end{subfigure}
     \hfill
     \begin{subfigure}[b]{0.45\textwidth}
         \centering
         \includegraphics[width=\textwidth]{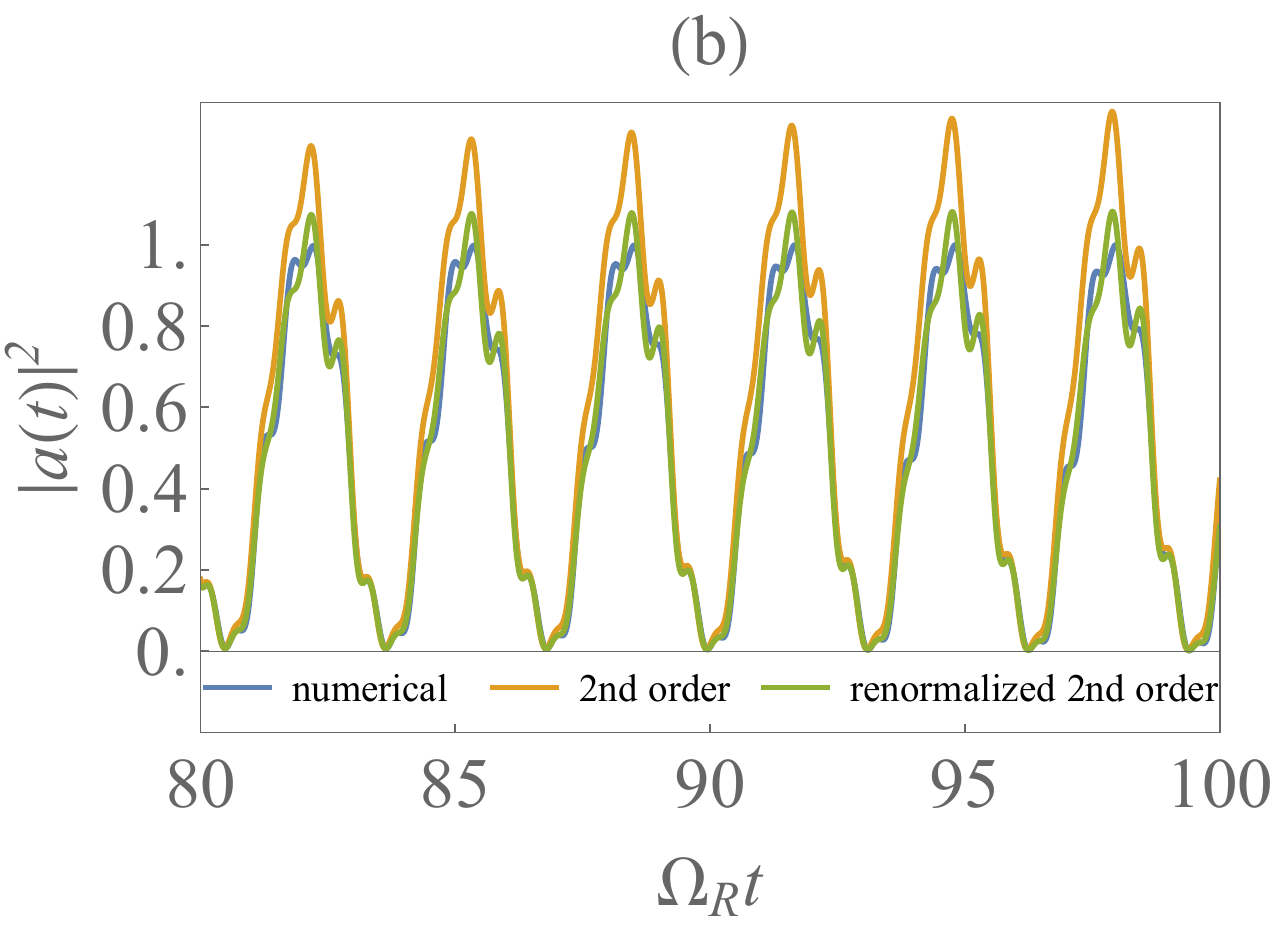}
         
     \end{subfigure}
     
     \begin{subfigure}[b]{0.45\textwidth}
         \centering
         \includegraphics[width=\textwidth]{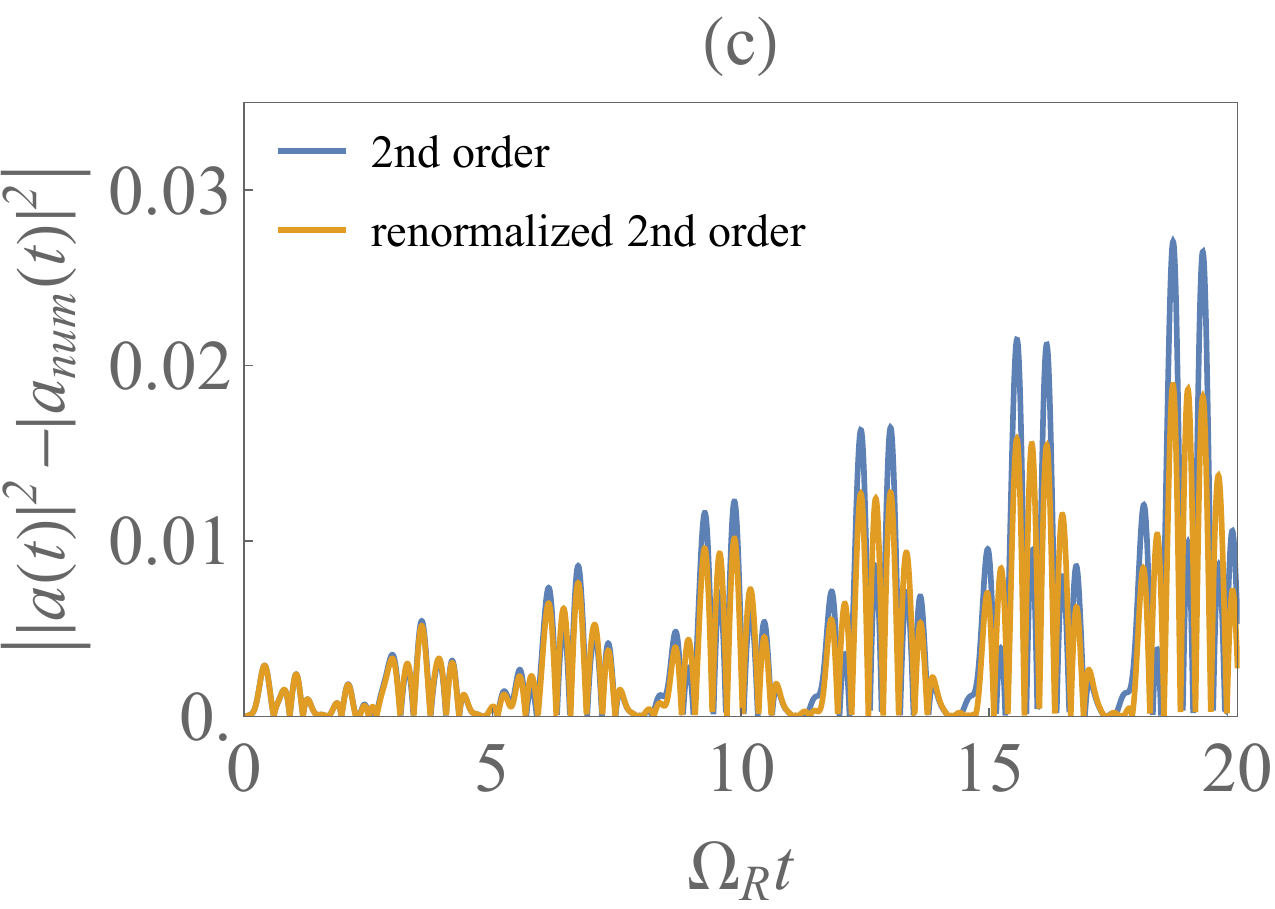}
        
     \end{subfigure}
     \hfill
     \begin{subfigure}[b]{0.45\textwidth}
         \centering
         \includegraphics[width=\textwidth]{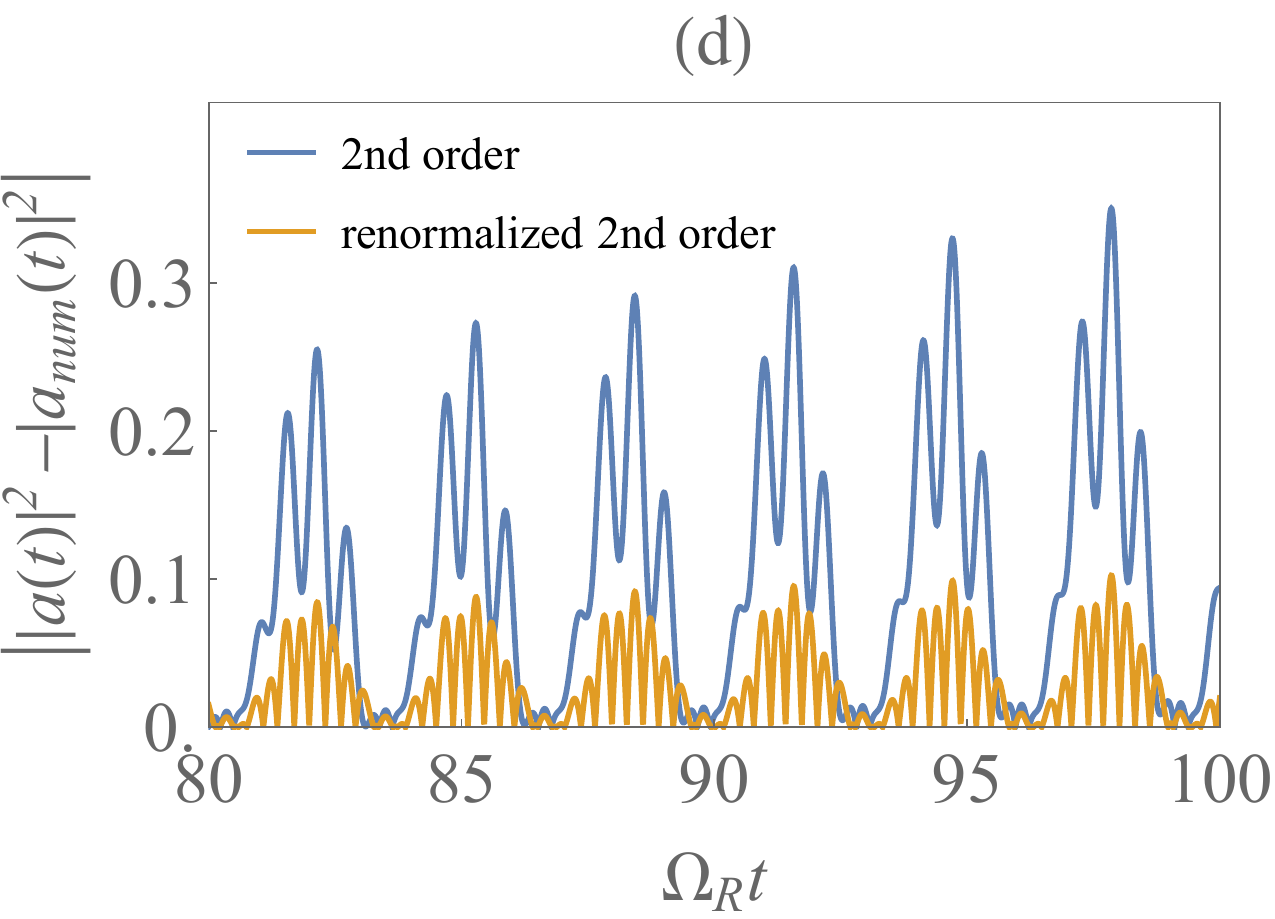}
         
     \end{subfigure}
\caption{\footnotesize Comparison of the renormalized multi-scale expansion Eq.~\eqref{eq:RWAfinal} with the two-scale expansion Eq.~\eqref{eq:rabi_expansion_two_scale_a2nd} for the Rabi model. Also shown are comparisons with numerical solutions to Eq.~\eqref{eq:eq_Rabi}. The probability $|a(t)|^2$ is plotted with $\delta_R=0$ and $\Delta_R=10$ for short times [(a) and (c)] and long times [(b) and (d)].}
\end{figure}

\section{Jaynes-Cummings model}

The Jaynes-Cummings model describes the interaction of a two-level atom with a single-mode quantized field. In this section, we investigate corrections to the RWA following the RG approach of Sec.~\ref{sec:Rabi}.
\subsection{Equations of motion}

We consider the Hamiltonian
\begin{eqnarray}
\begin{aligned}
\hat{H}&=\hat{H}_{A}+\hat{H}_{F}+\hat{H}_{AF} .
\end{aligned}
\end{eqnarray}
Here the atomic Hamiltonian $\hat{H}_{A}$ is given by Eq.~\eqref{eq:hamiltonian_atom}.
The Hamiltonian of the field $\hat{H}_F$ is of the form
\begin{eqnarray}
    \hat{H}_F=\hbar\omega\hat{a}^{\dagger}\hat{a},
\end{eqnarray}
where $\hat{a}^\dag$ and $\hat{a}$ are creation and annihilation operators of a field mode with energy $\hbar\omega$ and we have neglected the zero-point energy.
The Hamiltonian governing the interaction between the atom and the field is given by
\begin{eqnarray}
\begin{aligned}
\hat{H}_{AF} =\hbar(\hat{\sigma}+\hat{\sigma}^{\dagger})(g\hat{a}+g^{*}\hat{a}^{\dagger}),
\end{aligned}
\end{eqnarray}
where the lowering operator $\hat{\sigma} = |0\rangle\langle 1|$ and the raising operator  $\hat{\sigma}^{\dagger}=|1\rangle \langle 0|$. The coupling constant $g$ characterizes the interaction between the atom and the field and is pure imaginary. 

The dynamics of the system is governed by the Schr{\"o}dinger equation. In the interaction picture, the state
$|\psi_I\rangle$ obeys Eq.~\eqref{eq:eq_Schrodinger}. The Hamiltonian in the interaction picture $\hat{H}_I$ can be expressed as
\begin{eqnarray}
\begin{aligned}
\hat{H}_I&=e^{i\hat{H}_0 t/\hbar}\hat{H}_{AF}e^{-i\hat{H}_0t/\hbar}\\
&=-i\hbar \Omega_{JC}(e^{i(\omega_{0}-\omega) t} \hat{\sigma}^{\dagger} \hat{a}+e^{-i(\omega_{0}+\omega) t}\hat{\sigma}\hat{a})+\text{h.c.} ,
\end{aligned}
\end{eqnarray}
where  $\hat{H}_0 = \hat{H}_A+\hat{H}_F$ and $\Omega_{JC} = ig$. The state of the system is of the form
\begin{eqnarray}
    |\psi_{I}(t)\rangle=\sum_{n=0}^{\infty} a_n(t)|0,n\rangle + b_n(t)|1,n\rangle ,
\end{eqnarray}
where $|m,n\rangle$, with $m=0,1$ and $n=0,1,2, \ldots$, are eigenstates of the Hamiltonian $\hat{H}_0$.
We note that $a_n(t)$ and $b_n(t)$ are the probability amplitudes that the atom is in its ground state and excited state, respectively, with $n$ photons in the field. The conservation of probability is expressed by the relation
\begin{eqnarray}
\sum_{n=0}^\infty |a_n(t)|^2 + |b_n(t)|^2 = 1 .
\end{eqnarray}
Making use of  the Schr{\"o}dinger equation \eqref{eq:eq_Schrodinger}, we find that the coefficients $a_n(t)$ and $b_n(t)$ obey the equations of motion
\begin{eqnarray}
    \begin{aligned}
    \dot{a}_{n}(t)&=\Omega_{JC}\left(\sqrt{n}e^{-i(\omega_{0}-\omega) t}b_{n-1}(t)-\sqrt{n+1}e^{-i(\omega_{0}+\omega) t}b_{n+1}(t)\right),\\
    \dot{b}_{n}(t)&=\Omega_{JC}\left(\sqrt{n}e^{i(\omega_{0}+\omega) t}a_{n-1}(t)-\sqrt{n+1}e^{i(\omega_{0}-\omega) t}a_{n+1}(t)\right).
    \end{aligned}\label{eq:jc_eq_original}
\end{eqnarray}
For convenience, we rescale the time $t$ to $\Omega_{JC}t$, so that Eq.~\eqref{eq:jc_eq_original} becomes in dimensionless form
\begin{eqnarray}
    \begin{aligned}
    \dot{a}_{n}(t)&=\sqrt{n}e^{-i\delta_{JC} t}b_{n-1}(t)-\sqrt{n+1}e^{-i\Delta_{JC}t}b_{n+1}(t),\\
    \dot{b}_{n}(t)&=\sqrt{n}e^{i\Delta_{JC} t}a_{n-1}(t)-\sqrt{n+1}e^{i\delta_{JC} t}a_{n+1}(t),
    \end{aligned}
\label{eq:jc_eq_rescale}
\end{eqnarray}
where 
\begin{eqnarray}
    \delta_{JC} = \frac{\omega_0-\omega}{\Omega_{JC}}, \quad  \Delta_{JC} = \frac{\omega_0+\omega}{\Omega_{JC}}.
\end{eqnarray}

\subsection{Rotating wave approximation}
\label{sec:level2}
Following Sec.~\ref{sec:RWA}, we make the RWA by neglecting the fast rotating terms $e^{\pm i\Delta_{JC} t}$
in Eq.~\eqref{eq:jc_eq_rescale}. We find that on resonance, where $\delta_{JC}=0$, Eq.~\eqref{eq:jc_eq_rescale} becomes
\begin{eqnarray}
    \begin{aligned}
    \dot{a}_{n}=\sqrt{n}b_{n-1},\\
    \dot{b}_{n-1}=-\sqrt{n}a_{n} .
    \end{aligned}\label{eq:jc_rwa_eq}
\end{eqnarray}
We assume that the atom is initially in its ground state and one photon is present in the field. This corresponds to the following initial conditions on the amplitudes $a_n$ and $b_n$,
\begin{eqnarray}
    \begin{aligned}
    a_{1}(0)=1,\\
    a_0(0)=a_2(0)=\cdots =0,\\
    b_0(0)=b_1(0)=b_2(0)\cdots =0.
    \end{aligned}\label{eq:jc_initial}
\end{eqnarray}
We then find that the solution to Eq.~\eqref{eq:jc_rwa_eq} is given by
\begin{eqnarray}
\label{eq:RWAJC}
    \begin{aligned}
    a_1(t)&=\cos t,\\
    b_0(t)&=-\sin t.
    \end{aligned}
\end{eqnarray}	
This result is compared to the numerical solution of Eq.~\eqref{eq:jc_eq_rescale} in Fig.~5, for the case where $\delta_{JC}=0$ and $\Delta_{JC}=10$. We note that the coupled system Eq.~\eqref{eq:jc_eq_rescale} must be truncated for numerical computation. That is, we require
\begin{eqnarray}
    a_{n}=0,\quad b_{n-1}=0 \quad \text{for} \quad n \geq n_0 .
\end{eqnarray}
We find that $n_0=15$ is sufficient to guarantee convergence with six digits of accuracy. Evidently, the RWA is accurate at short times ($\Omega_{JC} t\approx 10$). 

\begin{figure}[t]
     \centering
     \begin{subfigure}[b]{0.45\textwidth}
         \centering
         \includegraphics[width=\textwidth]{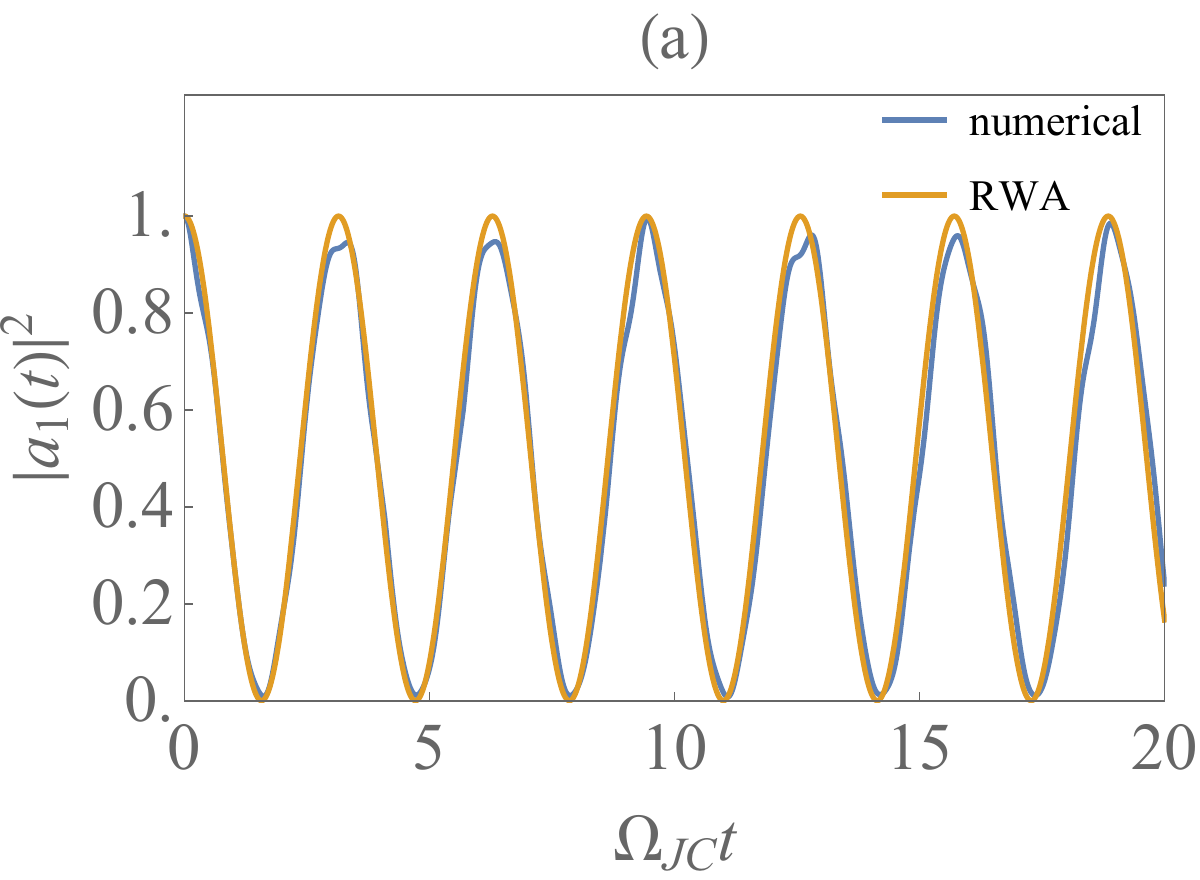}
        
     \end{subfigure}
     \hfill
     \begin{subfigure}[b]{0.45\textwidth}
         \centering
         \includegraphics[width=\textwidth]{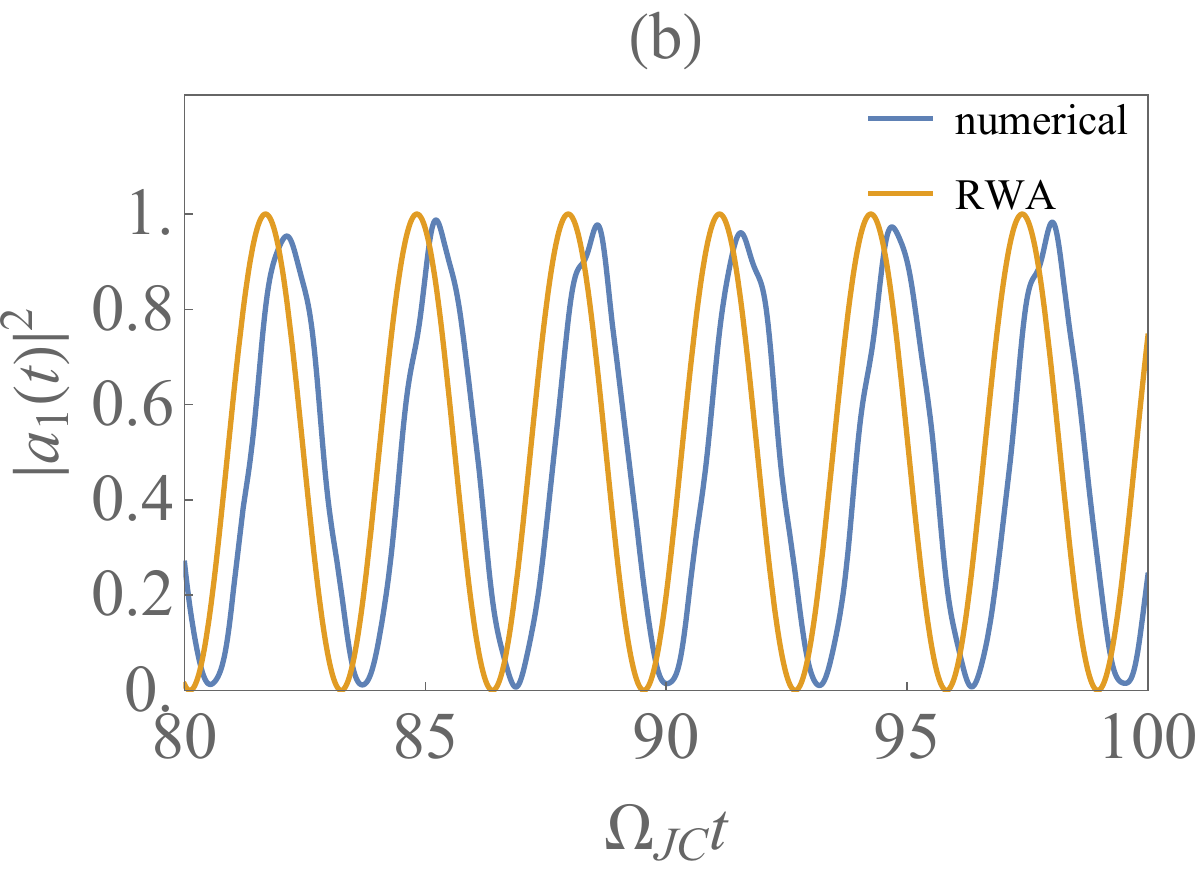}
         
     \end{subfigure}
\caption{\footnotesize Comparison of the RWA for the Jaynes-Cummings model with numerical solutions to the equations of motion Eq.~\eqref{eq:jc_eq_rescale}. The probability $|a_1(t)|^2$ is plotted with $\delta_R=0$ and $\Delta_R=50$ for (a) short times and (b) long times.}
\end{figure}

\subsection{Renormalized multi-scale expansion}

We now derive corrections to the RWA using renormalized perturbation theory, following the approach of
Sec.~\upperRomannumeral{3}. We begin by introducing a multi-scale expansion for $a_n$ and $b_n$:
\begin{eqnarray}
\begin{aligned}
     a_{n}^{\epsilon}(t_1, t_2) &= a_{n}(\epsilon t)=a_{n}^{(0)}(t_1,t_2)+\epsilon a_{n}^{(1)}(t_1,t_2)+\epsilon^{2}a_{n}^{(2)}(t_1,t_2)+\cdots,\\
     b_{n}^{\epsilon}(t_1, t_2) &= a_{n}(\epsilon t)=b_{n}^{(0)}(t_1,t_2)+\epsilon b_{n}^{(1)}(t_1,t_2)+\epsilon^{2}b_{n}^{(2)}(t_1,t_2)+\cdots,
\end{aligned}\label{eq:jc_expansion}
\end{eqnarray}
where $\epsilon = 1/\Delta_{JC}$, $t_1= t$ and $t_2= \epsilon t$. Eq.~\eqref{eq:jc_eq_rescale} then becomes
\begin{eqnarray}
    \begin{aligned}
    (\partial_{t_1}+\epsilon\partial_{t_2}){a}_{n}^{\epsilon}(t_1, t_2) &=\epsilon\bigl(\sqrt{n}b_{n-1}^{\epsilon}(t_1, t_2)-\sqrt{n+1}e^{-it}b_{n+1}^{\epsilon}(t_1, t_2)\bigr),\\
    (\partial{t_1}+\epsilon\partial_{t_2}){b}_{n}^{\epsilon}(t_1, t_2) &=\epsilon\bigl(\sqrt{n}e^{i t}a_{n-1}^{\epsilon}(t_1, t_2)-\sqrt{n+1}a_{n+1}^{\epsilon}(t_1, t_2)\bigr) ,
    \end{aligned}\label{eq:jc_remulti-scale_eq}
\end{eqnarray}
where we have made use of Eq.~\eqref{eq:derivative}.
The initial conditions Eq.~\eqref{eq:jc_initial} become
\begin{eqnarray}
\begin{aligned}
    a_{1}(0,0)=1,\\
    a_0(0,0)=a_2(0,0)=\cdots =0,\\
    b_0(0,0)=b_1(0,0)=b_2(0,0)\cdots =0.
    \end{aligned}\label{eq:jc_remulti-scale_initial}
\end{eqnarray}
It follows that  $a_{2n}^{\epsilon}(t_1,t_2)$ and $b_{2n+1}^{\epsilon}(t_1,t_2)$ vanish for $n=0,1,2,\ldots \ .$ According to Eq.~\eqref{eq:jc_remulti-scale_eq}, the remaining terms form a coupled infinite system. Remarkably, the system can be decoupled within the two-scale expansion Eq.~\eqref{eq:jc_expansion} and thus Eq.~\eqref{eq:jc_remulti-scale_eq} can be solved without truncation. Separating terms of order $\epsilon^{m}$ 
in Eq.~\eqref{eq:jc_remulti-scale_eq} we obtain
\begin{eqnarray}
    \begin{aligned}
    &\partial_{t_1}{a}_{2n+1}^{(m)}(t_1, t_2)+\partial_{t_2}{a}_{2n+1}^{(m-1)}(t_1, t_2)=\sqrt{2n+1}b_{2n}^{(m-1)}(t_1, t_2)-\sqrt{2(n+1)}e^{-it}b_{2n+2}^{(m-1)}(t_1, t_2),\\
    &\partial_{t_1}{b}_{2n}^{(m)}(t_1, t_2)+\partial_{t_2}{b}_{2n}^{(m-1)}(t_1, t_2)=\sqrt{2n}e^{i t}a_{2n-1}^{(m-1)}(t_1, t_2)-\sqrt{2n+1}a_{2n+1}^{(m-1)}(t_1, t_2).
    \end{aligned}\label{eq:jc_eq_morder}
\end{eqnarray}
When $m=0$, the solution to Eq.\eqref{eq:jc_eq_morder} is
\begin{eqnarray}
    \begin{aligned}
    a_{2n+1}^{(0)}&=\alpha_{2n+1}^{(0)}(t_2),\\
    b_{2n}^{(0)}&=\beta_{2n}^{(0)}(t_2) , \quad
    n=0,1,2, \ldots \ .
    \end{aligned}
\end{eqnarray}
Here $\alpha_{2n+1}^{(0)}(t_2)$ and $\beta_{2n}^{(0)}(t_2)$ are functions of $t_2$, which will be chosen so that the secular terms in $a_{2n+1}^{(1)}(t_1,t_2)$ and $b_{2n}^{(1)}(t_1,t_2)$ can be removed. By solving Eq.~\eqref{eq:jc_remulti-scale_eq} to first order in $\epsilon$, we find that $a_{2n+1}^{(1)}(t_1,t_2)$ and $b_{2n}^{(1)}(t_1,t_2)$ does not contain secular terms in $t_1$ provided $\alpha_{2n+1}^{(0)}(t_2)$ and $\beta_{2n}^{0}(t_2)$   satisfy
\begin{eqnarray}
    \begin{aligned}
    -\dot{\alpha}_{2n+1}^{(0)}(t_2)+\sqrt{2n}\beta_{2n}^{(0)}(t_2)&=0,\\
    \sqrt{2(n+1)}\alpha_{2n+1}^{(0)}(t_2)+\dot{\beta}_{2n}^{(0)}(t_2)&=0, \quad n=0,1,2,\ldots \ .
    \end{aligned}\label{eq:jc_eq_alpha_beta}
\end{eqnarray}
The solution to Eq.~\eqref{eq:jc_eq_alpha_beta} that obeys the initial conditions Eq.~\eqref{eq:jc_remulti-scale_initial} is
\begin{eqnarray}
\begin{aligned}
\alpha_{1}^{(0)}(t_2)=\cos t_2,\\
\beta_{0}^{(0)}(t_2)=-\sin t_2.\\
\end{aligned}
\end{eqnarray}
Therefore, we have
\begin{eqnarray}
    \begin{aligned}
    a_{1}^{(0)}(t_1,t_2)&=\cos t_2,\\
    a_{3}^{(0)}(t_1,t_,2)=a_{5}^{0}(t_1,t_2)=\cdots&=0,\\
    b_{0}^{(0)}(t_1,t_2)&=-\sin t_2,\\
    b_{2}^{(0)}(t_1,t_,2)=b_{4}^{0}(t_1,t_2)=\cdots&=0 \ ,
    \end{aligned}
\end{eqnarray}
which coincides with the RWA Eq.~\eqref{eq:RWAJC}.

Higher order terms, which lead to corrections to the RWA,  can be calculated in a similar manner. 
We find that
\begin{eqnarray}
    \begin{aligned}
    a_{1}^{(1)}(t_1,t_2)&=i(t_2\cos t_2+\sin t_2),\\
    a_{1}^{(2)}(t_1,t_2)&=\frac{1}{2}\bigl(-(4+t_2^{2})\cos t_2-t_2\sin t_2\bigr)+2e^{-i t_1}\cos\sqrt{3}t_2,\\
    a_{1}^{(3)}(t_1,t_2)&=-\frac{i}{6}\bigl((3t_2+t_2^{3})\cos t_2-15\sin t_2\bigr)+2ie^{-i t_1}(t_2\cos\sqrt{3}t_2+\sqrt{3}\sin\sqrt{3}t_2).
    \end{aligned}
\end{eqnarray}
In the above, we note the presence of secular terms in $t_2$. We then have
\begin{eqnarray}
    \begin{aligned}
       a_1^{\epsilon}(t_1,t_2)&= a_1^{(0)}(t_1,t_2)+\epsilon a_1^{(1)}(t_1,t_2)+\epsilon^{2} a_1^{(2)}(t_1,t_2)+\epsilon^{3} a_1^{(3)}(t_1,t_2)+O(\epsilon ^{4})\\
       &=\bigl(\cos t_2 +i\epsilon t_2\cos t_2 -\frac{1}{2}\epsilon^2(t_2^{2}\cos t_2+t_2\sin t_2)-\frac{i}{6}\epsilon^3(3t_2+t_2^{3})\cos t_2 \bigr)\\
       & \quad +2\epsilon^{2}e^{-it_1}(\cos\sqrt{3}t_2+i\epsilon t_2\cos\sqrt{3}t_2)+i\epsilon \sin t_2-2\epsilon^{2}\cos t_2\\
       & \quad +i\epsilon^{3}(2\sqrt{3}e^{-it_1}\sin\sqrt{3}t_2+\frac{5}{2}\sin t_2)+O(\epsilon  ^{4}).
    \end{aligned}\label{eq:jc_expression_a1}
\end{eqnarray}
Here, for the purpose of renormalization, we have grouped the secular terms with no dependence on $t_1$ together with the zeroth order non-secular term $\cos t_2$, because the latter is also independent of $t_1$. In addition, secular terms containing $e^{-it_1}$ were grouped with the second order non-secular term $2e^{-it_1}\cos\sqrt{3}t_2$. This rearrangement yields two groups to be renormalized:
\begin{eqnarray}
\begin{aligned}
    A_{\epsilon}^{1}(t_2) &=\cos t_2 +i\epsilon t_2\cos t_2 -\frac{1}{2}\epsilon^2(t_2^{2}\cos t_2+t_2\sin t_2)-\frac{i}{6}\epsilon^3(3t_2+t_2^{3})\cos t_2+O(\epsilon^{4}),\\
    A_{\epsilon}^{2}(t_2) &=\cos\sqrt{3}t_2+i\epsilon t_2\cos\sqrt{3}t_2+O(\epsilon^{2}).
\end{aligned}\label{eq:jc_expression_A}
\end{eqnarray}
The second group $A_{\epsilon}^{2}(t_2)$ can be renormalized as
\begin{eqnarray}
    A_{\epsilon}^{2}(t_2)=\cos\sqrt{3}t_2 e^{i\epsilon t_2}.
    \label{eq:jc_A_easy_renormalize}
\end{eqnarray}
In contrast, the first group $A_{\epsilon}^{1}(t_2)$ can be renormalized by including the non-secular term $i\epsilon \sin t_2$. To this end we introduce a new group $\widetilde A_{\epsilon}^1(t_2)$ which is defined as
\begin{eqnarray}
\begin{aligned}
    \widetilde A_{\epsilon}^1(t_2)(t_2)&=A_{\epsilon}^{1}(t_2)+i\epsilon \sin t_2\\
    &=\cos t_2 +i\epsilon( t_2\cos t_2+\sin t_2) -\frac{1}{2}\epsilon^2(t_2^{2}\cos t_2+t_2\sin t_2)-\frac{i}{6}\epsilon^3(3t_2+t_2^{3})\cos t_2+O(\epsilon^{4}),
\end{aligned}
\end{eqnarray}
which can be renormalized as
\begin{eqnarray}
     \widetilde A_{\epsilon}^1(t_2)(t_2)=\frac{1+\epsilon}{2}e^{it_2(1+\epsilon-\frac{1}{2}\epsilon ^{2})}+\frac{1-\epsilon}{2}e^{-it_2(1-\epsilon-\frac{1}{2}\epsilon^{2})}.
     \label{eq:jc_adjusted_A}
\end{eqnarray}
Using Eqs.~\eqref{eq:jc_expression_a1}, \eqref{eq:jc_A_easy_renormalize} and \eqref{eq:jc_adjusted_A}, we find that the renormalized multi-scale expansion of $a_1^{\epsilon}(t_1,t_2)$  is given by
\begin{eqnarray}
\begin{aligned}
    a_1^{\epsilon,R}(t_1,t_2)&=\frac{1+\epsilon}{2}e^{it_2(1+\epsilon-\frac{1}{2}\epsilon ^{2})}+\frac{1-\epsilon}{2}e^{-it_2(1-\epsilon-\frac{1}{2}\epsilon^{2})}+2\epsilon^{2}e^{-it_1}\cos\sqrt{3}t_2 e^{i\epsilon t_2}\\
    &-2\epsilon^{2}\cos t_2+i\epsilon^{3}(2\sqrt{3}e^{-it_1}\sin\sqrt{3}t_2+\frac{5}{2}\sin t_2)+O(\epsilon  ^{4}) .
\end{aligned}
\label{eq:jc_a1_expansion_3rd_order}
\end{eqnarray}
The above provides corrections to the RWA at long times and is our main result for the Jaynes-Cummings model. 

The probability $|a_1^{R}(t)|^{2}=|a_1^{\epsilon,R}(t/\epsilon)|^{2}$ is plotted in Fig.~6 with $\epsilon=1/\Delta_R=0.1$. Also shown is the comparison with the numerical solution to Eq.~\eqref{eq:jc_eq_rescale}. We see that Eq.\eqref{eq:jc_a1_expansion_3rd_order} provides corrections to the RWA at times of the order of $O(1/\epsilon^{2})$.

\begin{figure}[t]
     \centering
     \begin{subfigure}[b]{0.45\textwidth}
         \centering
         \includegraphics[width=\textwidth]{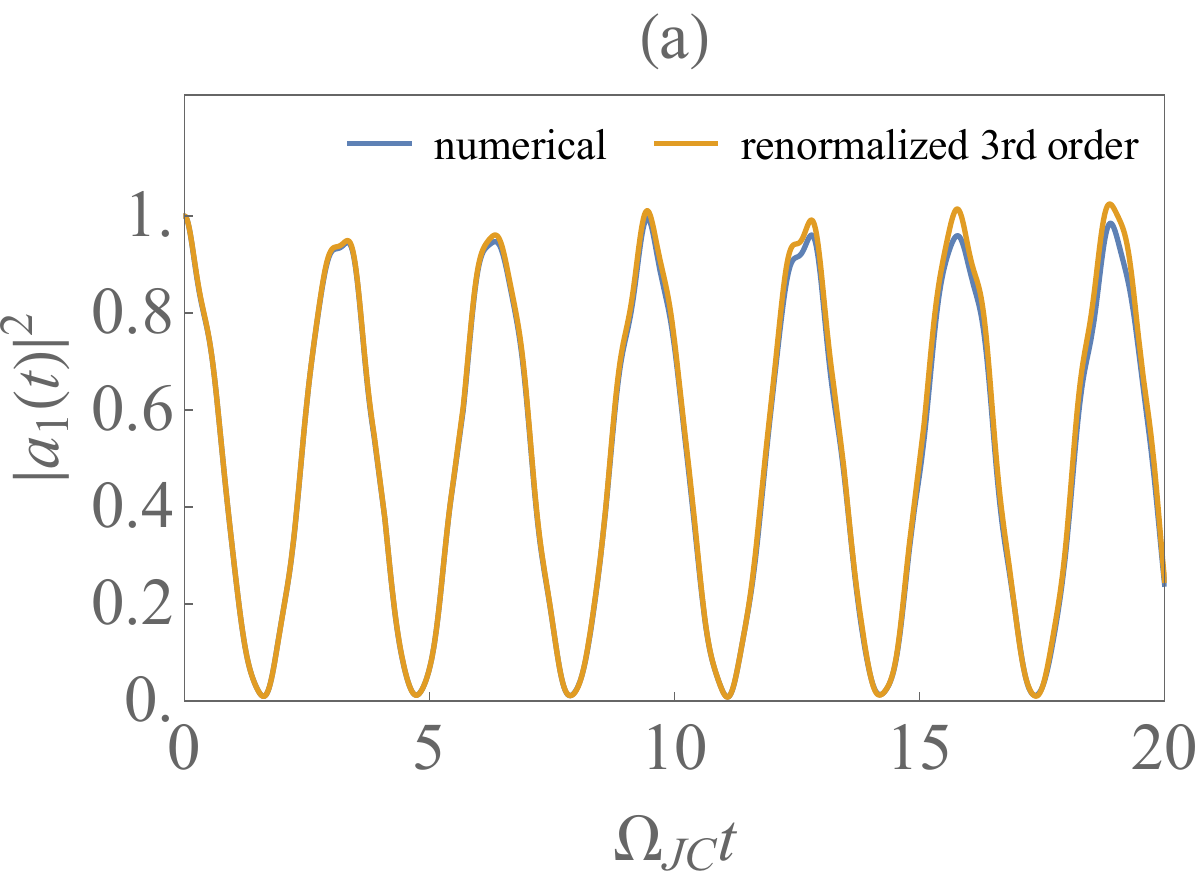}
        
     \end{subfigure}
     \hfill
     \begin{subfigure}[b]{0.45\textwidth}
         \centering
         \includegraphics[width=\textwidth]{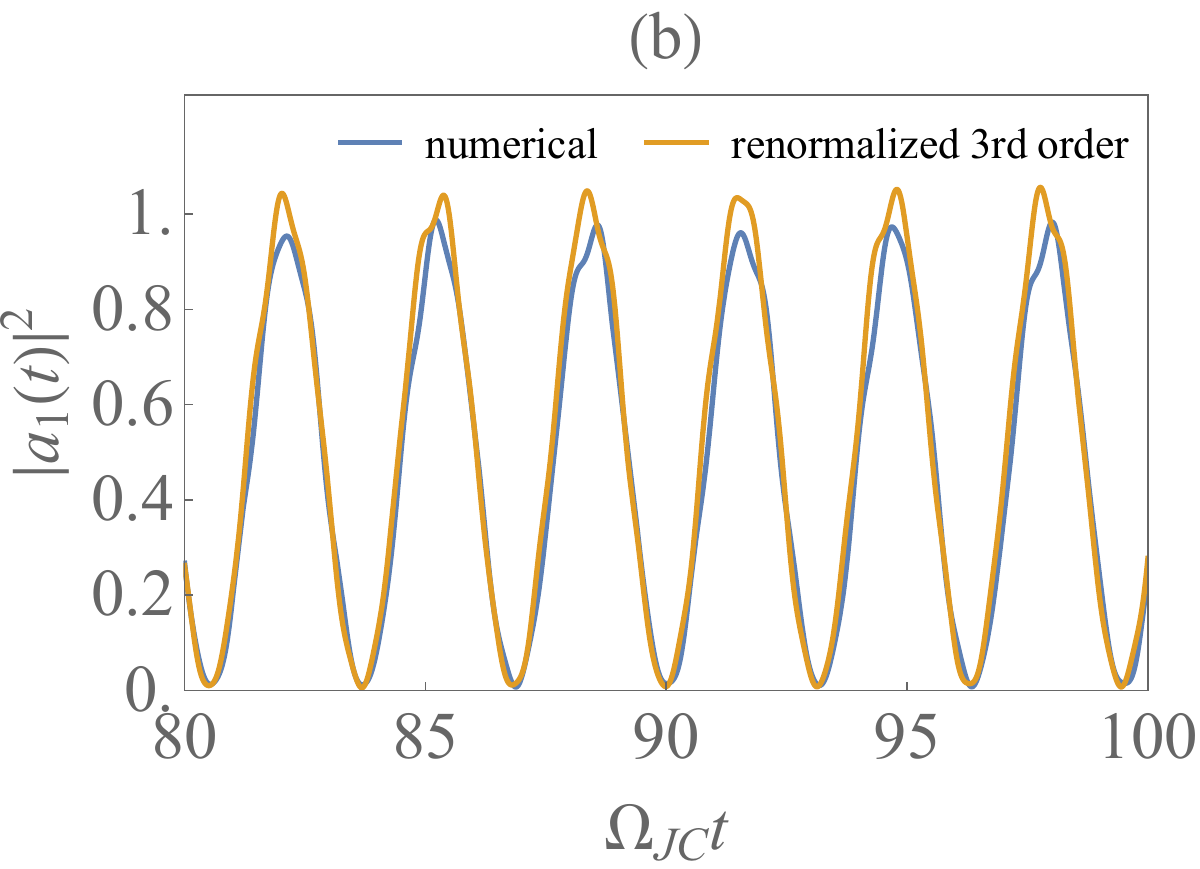}
         
     \end{subfigure}
     
     \begin{subfigure}[b]{0.45\textwidth}
         \centering
         \includegraphics[width=\textwidth]{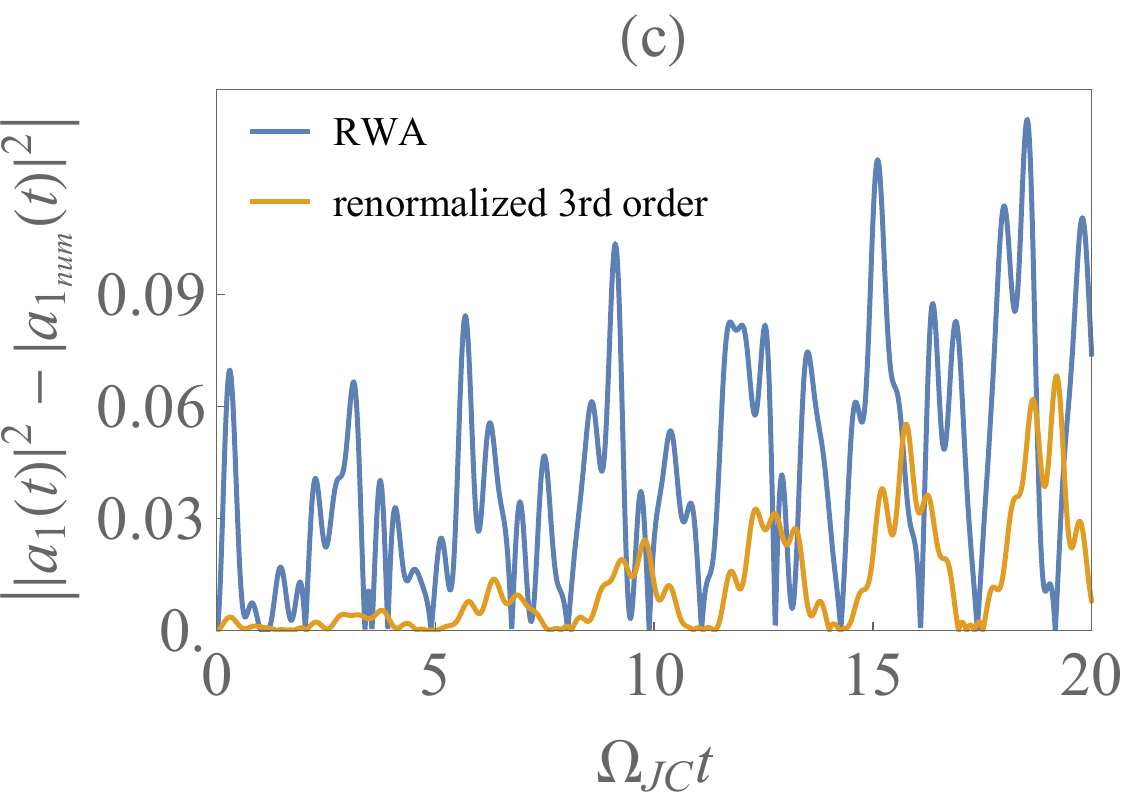}
        
     \end{subfigure}
     \hfill
     \begin{subfigure}[b]{0.45\textwidth}
         \centering
         \includegraphics[width=\textwidth]{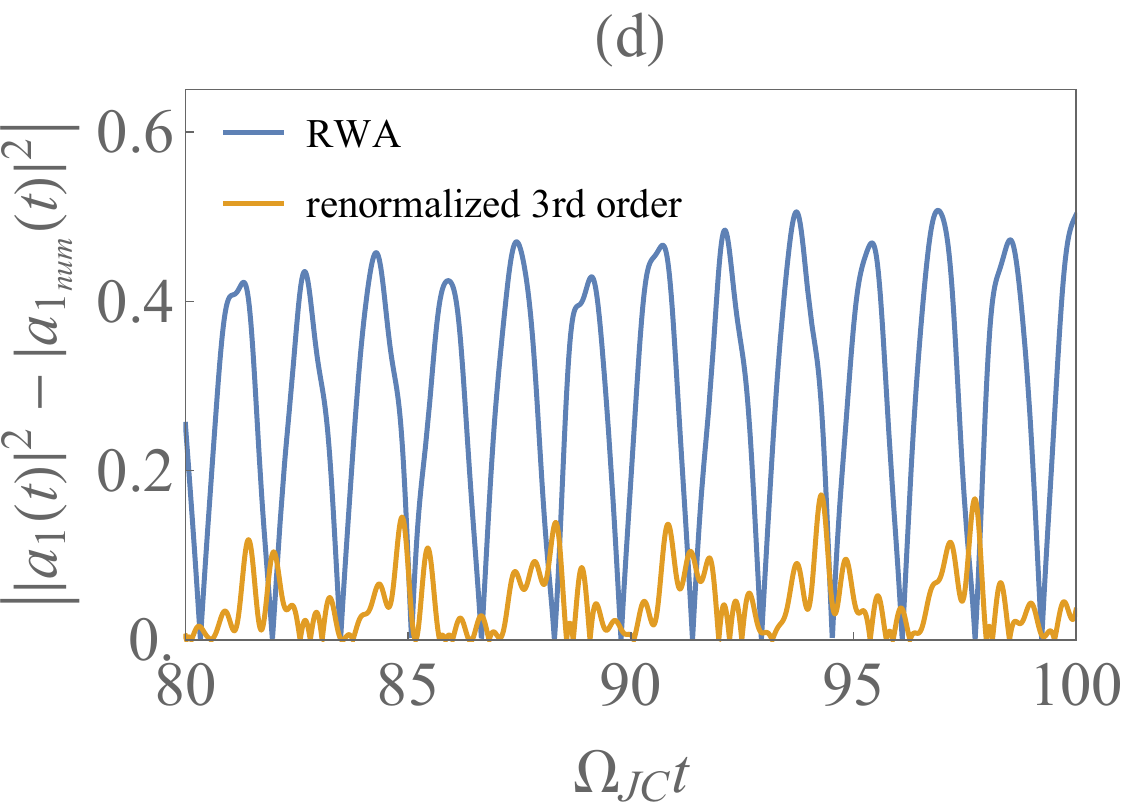}
         
     \end{subfigure}
\caption{\footnotesize Comparison of the renormalized multi-scale expansion Eq.~\eqref{eq:jc_a1_expansion_3rd_order} with the RWA Eq.~\eqref{eq:RWAJC} for the Jaynes-Cummings model. Also shown are comparisons with numerical solutions to Eq.~\eqref{eq:jc_eq_rescale}. The probability $|a_1(t)|^2$ is displayed with $\delta_R=0$ and $\Delta_R=10$ for short times [(a) and (c)] and long times [(b) and (d)].}
\end{figure}

\begin{figure}[t]
     \centering
          \includegraphics[width=.5\textwidth]{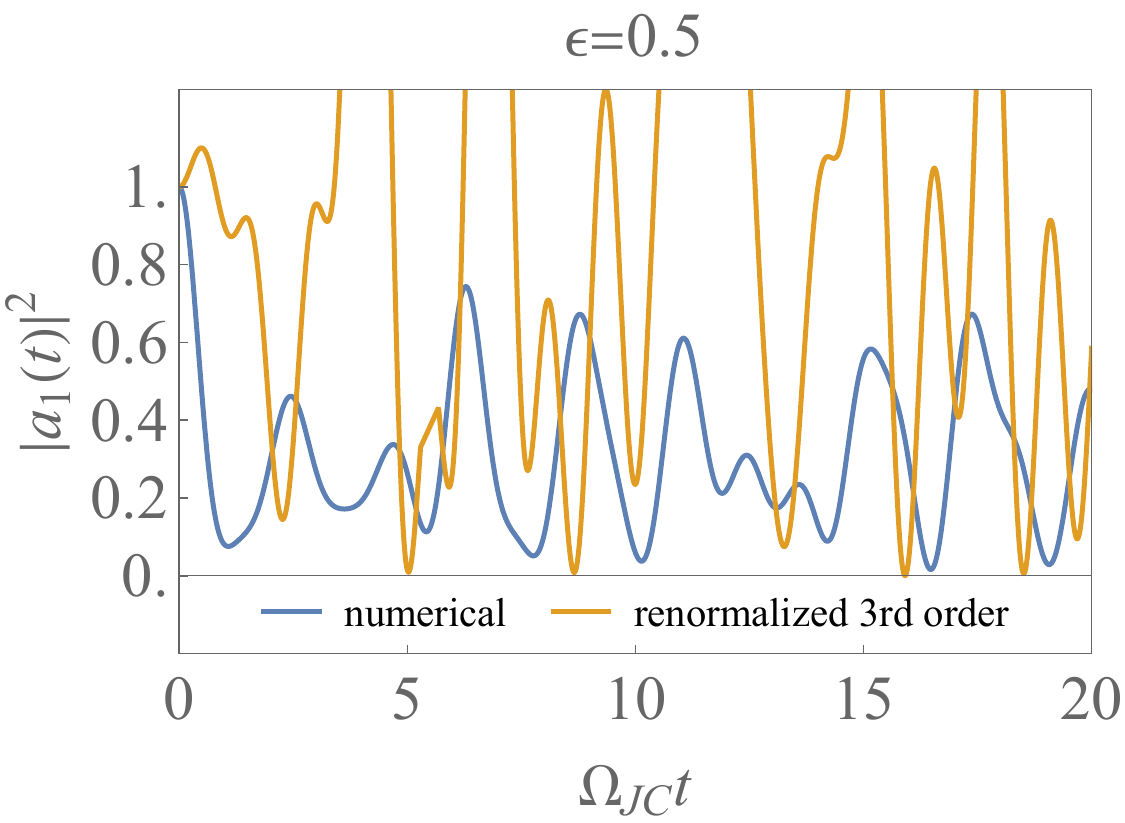}
    \caption{\footnotesize Comparison of the renormalized multi-scale expansion Eq.~\eqref{eq:jc_a1_expansion_3rd_order} with numerical solutions to the equations of motion Eq.~\eqref{eq:jc_eq_rescale} for the Jaynes-Cummings model. The probability $|a_1(t)|^2$ is shown with $\delta_R=0$ and $\Delta_R=2$.}
\end{figure}

\section{Discussion}

In this paper we have derived corrections to the RWA for the Rabi and Jaynes-Cummings models. Our results are obtained from renormalized perturbation theory applied to the equations of motion for a single-excitation state. In this manner, we obtain corrections to the RWA that are free of secular terms to finite order in perturbation theory.

We now comment on the possible relevance of our results to experiment. The Jaynes-Cummings model is often applied to experiments in cavity and circuit quantum electrodynamics. In this setting, the quantity $\epsilon=1/\Delta_{JC}$ is a measure of the atom-field coupling, which in practice ranges from $10^{-6} \le \epsilon \le 10^{-2}$~\cite{Sillanpa_2007,Majer_2007,Fink_2008,Deppe_2008}. Thus the results illustrated in Fig.~6, where $\epsilon=10^{-1}$, are a stringent test of the proposed renormalized perturbation expansion Eq.~\eqref{eq:jc_a1_expansion_3rd_order}.
To further investigate the effect of strong coupling, we set $\epsilon=0.5$. As shown in Fig.~7, it can be seen that Eq.~\eqref{eq:jc_a1_expansion_3rd_order} breaks down in this instance. This finding is consistent with the chaotic dynamics of the system Eq.~\eqref{eq:jc_eq_rescale}~\cite{Crisp_1991,Bonci_1991,Emary_2003}.

Several topics for further research are apparent. It would be of interest to explore corrections to the RWA for a continuum of field modes. This would entail the study of single- and multi-atom problems and the dynamics of superadiance. In addition, the investigation of two-photon problems is of considerable current interest, especially effects related to entanglement in many-body systems.

\appendix
\section{Ricatti equation for the Rabi model}
Here we present an alternative approach to finding corrections to the RWA for the Rabi model.
We begin by recalling Eq.~\eqref{eq:eq_rabi_rwa_resonance} and define
\begin{eqnarray}
    u(t)= \frac{b(t)}{a(t)} .
    \label{eq:ricatti_def_u}
\end{eqnarray}
We find that $u$ obeys the Ricatti equation
\begin{eqnarray}
    i\dot{u}(t)+(1+e^{-i\Delta_R t})u^2(t)-(1+e^{i\Delta_R t})=0 .
    \label{eq:ricatti_eq}
\end{eqnarray}
Recall the conservation of probability,
\begin{eqnarray}
    |a(t)|^2+|b(t)|^2=1.
\end{eqnarray}
Thus the probabilities $|a(t)|^2$ and $|b(t)|^2$ can be calculated from $u(t)$ using the relations
\begin{eqnarray}
    |a(t)|^{2}=\frac{1}{1+|u(t)|^{2}} , \quad |b(t)|^{2}=\frac{|u(t)|^2}{1+|u(t)|^2} .
\end{eqnarray}

The renormalized multi-scale expansion method developed in Section \upperRomannumeral{3} can be used to derive an asymptotic to Eq.\eqref{eq:ricatti_eq}. Here we  present the result to second order:
\begin{eqnarray}
\begin{aligned}
    &u_{R}(t)=u_0(t)+\epsilon \left(-u^{2}_{0}(t)e^{-it/\epsilon}-e^{it/\epsilon}+\frac{1+2e^{2it(1+\frac{\epsilon}{2})}+e^{4it}}{(1+e^{2it})^2}\right)\\
&+\epsilon^{2}\left(2iu_0(t)\left(\dot{u}_{0}(t)-iu^2_0(t)+i\right)e^{-it/\epsilon}-2u_0(t)e^{it/\epsilon}+u^{3}_{0}(t)e^{-2it/\epsilon}
+\frac{3(1-e^{4it})}{2(1+e^{2it})^2}\right)\\
&+O(\epsilon^{3}),
\end{aligned}  \label{eq:ricatti_expansion_solution}
\end{eqnarray}
where $\epsilon = 1/\Delta_R$  and
\begin{eqnarray}
    u_0(t)=\frac{1-e^{2it}}{1+e^{2it}} .
\end{eqnarray}
Fig.~8 shows numerical calculations of the probability $|a(t)|^2$  compared with the RWA and Eq.~\eqref{eq:ricatti_expansion_solution} for $\delta_R=0$ and $\Delta_R=10$.

\begin{figure}[t]
     \centering
         \includegraphics[width=.5\textwidth]{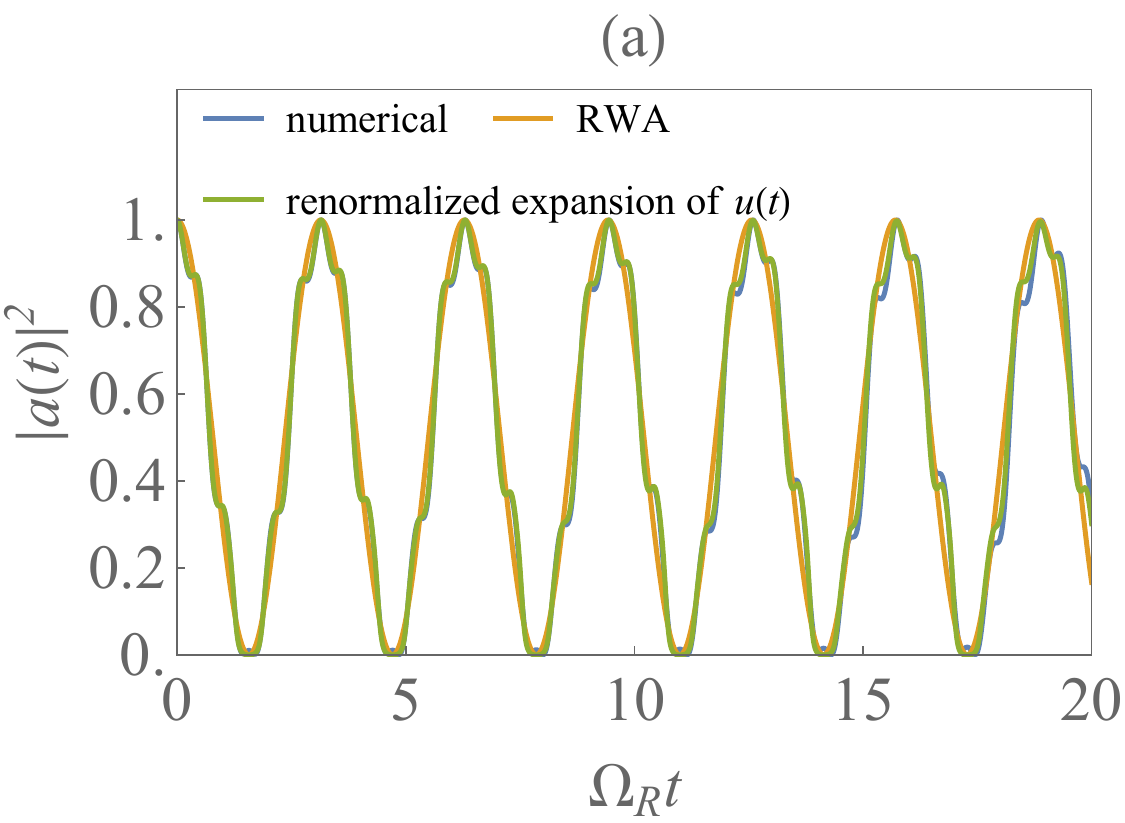}
\caption{\footnotesize Comparison of the RWA for the Rabi model with the two-scale expansion Eq.~\eqref{eq:ricatti_expansion_solution} and numerical solutions to the equations of motion Eq.~\eqref{eq:eq_Rabi}. The probability $|a(t)|^2$ is plotted with $\delta_R=0$ and $\Delta_R=10$.}
\end{figure}

\section*{Acknowledgments}
This work was supported in part by the NSF grant DMS-1912821 and the AFOSR grant FA9550-19-1-0320.

\section*{Author Declarations}

\subsection*{Conflict of Interest}

The authors have no conflicts to disclose.

\section*{Data Sharing}

Data sharing is not applicable to this article since no data were created in this study.

\end{document}